\documentclass[twocolumn]{aastex631}

\begin{document}

\title{The FAST Ursa Major supergroup H\,{\footnotesize I} survey (FUMaS): catalog and H\,{\footnotesize I} mass function}

\correspondingauthor{Haiyang Yu, Ming Zhu}
\email{hyyu@nao.cas.cn, mz@nao.cas.cn}

\author[0000-0001-9838-7159]{Haiyang Yu}
\affiliation{National Astronomical Observatories, Chinese Academy of Sciences, A20 Datun Road, Beijing 100101, China}
\affiliation{University of Chinese Academy of Sciences, Beijing 100049, China}
\affiliation{CAS Key Laboratory of FAST, National FAST, National Astronomical Observatories, Chinese Academy of Sciences, Beijing 100101, China}

\author[0000-0001-6083-956X]{Ming Zhu}
\affiliation{National Astronomical Observatories, Chinese Academy of Sciences, A20 Datun Road, Beijing 100101, China}
\affiliation{University of Chinese Academy of Sciences, Beijing 100049, China}
\affiliation{CAS Key Laboratory of FAST, National FAST, National Astronomical Observatories, Chinese Academy of Sciences, Beijing 100101, China}
\affiliation{Guizhou Radio Astronomical Observatory, Guizhou University, Guiyang 550000, People's Republic of China}

\author[0000-0002-4428-3183]{Chuan-Peng Zhang}
\affiliation{National Astronomical Observatories, Chinese Academy of Sciences, A20 Datun Road, Beijing 100101, China}
\affiliation{CAS Key Laboratory of FAST, National FAST, National Astronomical Observatories, Chinese Academy of Sciences, Beijing 100101, China}
\affiliation{Guizhou Radio Astronomical Observatory, Guizhou University, Guiyang 550000, People's Republic of China}

\author[0000-0002-5387-7952]{Peng Jiang}
\affiliation{National Astronomical Observatories, Chinese Academy of Sciences, A20 Datun Road, Beijing 100101, China}
\affiliation{CAS Key Laboratory of FAST, National FAST, National Astronomical Observatories, Chinese Academy of Sciences, Beijing 100101, China}
\affiliation{Guizhou Radio Astronomical Observatory, Guizhou University, Guiyang 550000, People's Republic of China}

\author{Jin-Long Xu}
\affiliation{National Astronomical Observatories, Chinese Academy of Sciences, A20 Datun Road, Beijing 100101, China}
\affiliation{CAS Key Laboratory of FAST, National FAST, National Astronomical Observatories, Chinese Academy of Sciences, Beijing 100101, China}
\affiliation{Guizhou Radio Astronomical Observatory, Guizhou University, Guiyang 550000, People's Republic of China}



\begin{abstract}

Using the Five-hundred-meter Aperture Spherical radio Telescope (FAST), we have performed an Ursa Major supergroup H\,{\footnotesize I} Survey (FUMaS) covering the entire UMa region centered at RA=11$^h$59$^m$28$^s$.3, DEC=49\degr05\arcmin18\arcsec with a radius of 7.5\degr.
We have obtained the most complete catalog of H\,{\footnotesize I} sources in the UMa supergroup, containing 178 H\,{\footnotesize I} sources with velocities in the range 625-1213.4 km~s$^{-1}$ and masses in the range 10$^{6.0}$-10$^{10.1}$ M$_{\sun}$ assuming a distance of 17.4 Mpc.
Among them, 55 H\,{\footnotesize I} sources were detected for the first time, of which 32 do not have known optical redshifts. 
For these 32 sources, we have searched the DESI Legacy Surveys and found optical counterparts for 25 of them (with optical images, but no redshifts), with the remaining 7 sources to be pure H\,{\footnotesize I} clouds without an optical counterpart.
We detected H\,{\footnotesize I} distributions in some interacting systems and discussed four small groups in detail.
We computed the H\,{\footnotesize I}MF of the UMa supergroup using the 1/V$_\mathrm{max}$ method and fitting it with the non-linear least squares (NLLS) and modified maximum likelihood (MML) methods.
We obtained the following H\,{\footnotesize I}MF parameters: log$_{10}$($\phi_*$/Mpc$^{-3}$) = -0.78 $\pm$ 0.15, $\alpha$ = -1.05 $\pm$ 0.07 and log$_{10}$($M_*$/$M_{\sun}$) = 9.87 $\pm$ 0.19 for the NLLS method, and log$_{10}$($\phi_*$/Mpc$^{-3}$) = -0.70 $\pm$ 0.11, $\alpha$ = -1.05 $\pm$ 0.05 and log$_{10}$($M_*$/$M_{\sun}$) = 9.77 $\pm$ 0.13 for the MML method.
This result is similar to that derived from the VLA blind survey, but the slope is steeper because we detected more low-mass galaxies.
The slope is flatter than that of the global H\,{\footnotesize I}MF, which agrees with the theoretical prediction that galaxies in high-density regions are stripped of gas due to interactions.

\end{abstract}



\section{Introduction} 
\label{sec:intro}

The Lambda Cold Dark Matter ($\Lambda$CDM) model of the Universe, validated by many observations, depicts the hierarchical formation of structures of the Universe \citep{White1978}. 
In this theoretical framework, small, Gaussian fluctuations are influenced by gravity to form dark matter halos through mergers and accretions. 
In dark matter halos, the surrounding baryons gradually cool and accrete to form galaxies, groups, clusters, filaments, and other large-scale structures (LSS). 
Although the $\Lambda$CDM model has led to a better understanding of the formation of LSS, the state of the early Universe, and the cosmic abundance of matter and energy, there are still some observational discrepancies that need to be resolved \citep{Bull2016}. 
For example, the slope at the low-mass end of the dark-matter halo mass function \citep{Jenkins2001} is much steeper than that of the galaxy luminosity function \citep[LF,][]{Blanton2001,Blanton2003} and the neutral hydrogen mass function \citep[H\,{\footnotesize I}MF,][]{Zwaan2005,Jones2018}, i.e., there are a large number of predicted low-mass dark-matter halos that have not been observed. 
To address this discrepancy, there are two directions of theory, one that modifies the nature of dark matter to suppress the formation of low-mass halos or increases their destruction \citep{Spergel2000,Kravtsov2004}, and another that suppresses star formation in low-mass halos so that they are not observed \citep{Verde2002,Benitez-Llambay2020}.

H\,{\footnotesize I} is an important component of galaxies, and its 21 cm emission line can provide many additional important parameters for galactic studies. 
A growing number of H\,{\footnotesize I} blind surveys have been conducted to explore the distributional properties of H\,{\footnotesize I} sources in the Universe. 
In contrast to optical and infrared surveys, H\,{\footnotesize I} blind surveys can detect gas-rich, low luminosity, or low surface brightness galaxies that cannot be optically detected. 
This is a good complement to the detection of low-mass galaxies. 
H\,{\footnotesize I} surveys that have been carried out over large areas of the sky include the H\,{\footnotesize I} Parkes All Sky Survey (HIPASS) and the Arecibo Legacy Fast ALFA (ALFALFA) \citep{Giovanelli2015}.
HIPASS covers the entire southern sky area south of the declination +2\degr and identified 4315 H\,{\footnotesize I} sources. 
It has a root mean square (RMS) noise of 13 mJy~beam$^{-1}$ and a velocity range of -1280 to 127000 km~s$^{-1}$ \citep{Barnes2001, Meyer2004}. 
ALFALFA has a higher sensitivity, detecting ~31500 H\,{\footnotesize I} sources with z$<$0.06 over an observation range of ~7000 square degrees \citep{Giovanelli2005, Haynes2018}.

The H\,{\footnotesize I}MF \citep{Zwaan2005, Martin2010, Jones2018, Ma2024} represents the distribution of the number of galaxies in different mass intervals, and it follows the Schechter function \citep{Schechter1976}, whose characteristic parameters are the slope $\alpha$ describing the low-mass end and the characteristic 'knee' mass $M_*$ at the high-mass end, respectively.
The first attempt to derive the H\,{\footnotesize I}MF was from Arecibo's blind H\,{\footnotesize I} survey.
\citet{Zwaan1997} used data from the Arecibo H\,{\footnotesize I} Strip Survey, which covered 65 square degrees and contained 66 detections, and measured the slope of the H\,{\footnotesize I}MF to be about -1.2.
\citet{Rosenberg2002} used the Arecibo Dual-beam Survey data covering an area of approximately 430 square degrees and measured a steeper slope of approximately -1.5.
As more and more large surveys are carried out, the overall distribution of H\,{\footnotesize I} in the universe becomes more and more clear.  
The low-mass slope parameter calculated by HIPASS using 4315 detections is -1.37, and the 'knee' mass is 9.80 \citep{Zwaan2005}.
The data from 100\% ALFALFA give a slightly different result, with a low mass slope of -1.25 and a ‘knee’ mass of 9.94 \citep{Jones2018}. 
The current MIGHTEE Early Science data \citep{Maddox2021}, based on the MeerKAT radio interferometer, give the first H\,{\footnotesize I}MF based on two different methods: 1/V$_\mathrm{max}$ and Modified Maximum Likelihood (MML), whose parameters are $\alpha$ = -1.29, $M_*$ = 10.07 for 1/V$_\mathrm{max}$ and $\alpha$ = -1.44, $M_*$ = 10.22 for MML, respectively \citep{Ponomareva2023}.
There is some variability in the present H\,{\footnotesize I}MF, which may be due to differences in the observing regions, as the distribution of H\,{\footnotesize I} is correlated with the environment and morphology.

\citet{Springob2005} optically selected a large number of spiral galaxies with morphology later than S0a for the study of the effects of morphology and environment on the H\,{\footnotesize I}MF. 
The results show a significant dependence of the H\,{\footnotesize I}MF on morphology, with galaxies with later morphology having steeper low-mass slopes of the H\,{\footnotesize I}MF. 
Higher density regions have a flatter H\,{\footnotesize I}MF low-mass slope and lower 'knee' mass, but the variability is insignificant.
\citet{Zwaan2003} obtained similar results with the H\,{\footnotesize I}MF of the 1000 brightest galaxies from HIPASS, with late-type galaxies having steeper slopes.
\citet{Moorman2014} selected about 7300 galaxies from ALFALFA and classified them as void galaxies (low density) versus wall galaxies (high density), with galaxies in the void regions having lower H\,{\footnotesize I} mass.
\citet{Jones2016} investigated the dependence of the H\,{\footnotesize I}MF on the environment using 70\% of the ALFALFA data. 
Their work delineated the surrounding density of galaxies based on the Sloan Digital Sky Survey (SDSS) and 2MASS Redshift Survey (2MRS) catalogs.
They found a dependence of the 'knee' mass on the environment, which grew from 9.81 to 10.00 as the SDSS galaxy density increased.
In contrast, the density delineation based on the 2MRS catalogs did not find any dependence on the environment. 
In addition, the dependence of the low-mass slope on the environment was not found in this work.
The H\,{\footnotesize I}MF for the Parkes H\,{\footnotesize I} Zone of Avoidance (HIZOA) survey shows that the low-mass slope steepens with decreasing density, while there is no clear trend for the ‘knee’ mass \citep{Said2019}.
In addition to the environmental classification of galaxies in large-scale blind surveys, direct observation of targeting specific environments is another way to study the environmental effects on the H\,{\footnotesize I}MF.
For observations of specific high-density groups and clusters of galaxies, the slopes of most of the H\,{\footnotesize I}MFs show a flatter slope \citep{Freeland2009, Pisano2011, Westmeier2017, Busekool2021} compared to the global H\,{\footnotesize I}MF, except for the Leo I group of galaxies, which has a slope of -1.41, but with a large error ($\pm 0.2$) \citep{Stierwalt2009}.
The environment has a strong correlation with H\,{\footnotesize I}MF, but the precise dependence remains to be explored.

The Ursa Major (UMa) region provides an excellent environment in which to study the influence of the environment on the H\,{\footnotesize I}MF.
With a distance of 17.4 Mpc \citep{Tully2012}, it is one of the three moderately sized regions in the Milky Way neighborhood.
However, UMa has received less attention than the Virgo and Fornax clusters because of its location at the junction of large-scale filamentary structures and the lack of any concentrating core.
Recent study suggests that the UMa's properties do not meet the criteria for a galaxy cluster, but it is still more complex than a group.
\citet{Wolfinger2016} redefined it as a supergroup rather than a cluster, in which groups are in an early stage of evolution, gravitationally bound to each other, and will form a larger structure and eventually merge with Virgo.
\citet{Tully1996} first defined the UMa region as a cluster within a circle centered at RA=11$^h$59$^m$28$^s$.3, DEC=49\degr05\arcmin18\arcsec, and with a radius of 7.5\degr.
They identified 79 members with a velocity range of 700-1210 km~s$^{-1}$ and a small velocity dispersion of 148 km~s$^{-1}$. 
\citet{Pak2014} identified 166 galaxies and studied their morphological distributions with spectroscopic data from SDSS-DR7 and the NASA/IPAC Extragalactic Database (NED). 
The result suggests that the UMa supergroup is dominated by late-type galaxies, but there are also a large number of early-type galaxies, especially in the dwarf region.
\citet{Verheijen2001} studied the H\,{\footnotesize I} properties and kinematics of 43 spiral galaxies in UMa with the Westerbork Synthesis Radio Telescope (WSRT).
\citet{Busekool2021} calculated the H\,{\footnotesize I}MF of the UMa supergroup using data from these WSRT observations as well as from a blind H\,{\footnotesize I} survey of a 16\% region of the UMa supergroup with the Very Large Array (VLA).
They found a low slope ($\alpha=-0.92$) in the H\,{\footnotesize I}MF, which indicates a lack of a gas-rich dwarf galaxies in the UMa supergroup.
The H\,{\footnotesize I} Jodrell All Sky Survey (HIJASS) also conducted a blind survey of the UMa region, but did not cover the complete circular region \citep{Wolfinger2013}. 
In addition, the work of HIPASS and ALFALFA did not cover the UMa region.
So, the study of the UMa supergroup lacks a complete catalog of H\,{\footnotesize I} sources.
The current FAST All Sky H\,{\footnotesize I} survey (FASHI) \citep{Zhang2024} using the Five-hundred-meter Aperture Spherical radio Telescope (FAST) can cover this region completely, and the high sensitivity of FAST can give us a new perspective on the H\,{\footnotesize I} distribution of the UMa supergroup. 
A number of H\,{\footnotesize I} source detections have been reported in the first data release of FASHI, but data for the UMa supergroup are not yet complete due to the schedule-filler observation strategy of the FASHI project.
This work was designed to complete the survey of the UMa supergroup by make additional observations and data analysis. 

In this paper, we present the catalog and H\,{\footnotesize I}MF resulting from the FAST UMa supergroup H\,{\footnotesize I} survey (FUMaS).
In Section \ref{sec:obs and dp}, we describe the observational setup of the project and data processing. 
Section \ref{sec:UMa catalog} describes the catalog of the UMa supergroup and compares them with existing catalogs. 
Section \ref{sec:discussion and results} discusses H\,{\footnotesize I} sources with optical counterparts without a redshift and small groups in the catalog.
The calculation of the H\,{\footnotesize I}MF is described in Section \ref{sec:HIMF}. 
It is summarized in Section \ref{sec:conclusion}.

\section{Observations and Data Processing}
\label{sec:obs and dp}

\subsection{Observations}
\label{sec:obs} 

The data used in this paper come from observations made by the FASHI project of FAST in 2021-2023 and additional observations in January 2024.
FAST is a single-dish radio telescope with an aperture of 500 m and an illuminating aperture of 300 m located in Guizhou, China \citep{Jiang2019, Jiang2020, Qian2020}. 
It has a zenith angle (ZA) of up to 40\degr, so it can observe over a range of declinations from -14.4 to 65.6\degr, and covers a frequency range of 1050-1450 MHz.
FASHI is a blind survey of the full sky area within which FAST can observe, and it is expected to detect more than 100000 H\,{\footnotesize I} sources over 22000 square degrees. 
The first release catalog is publicly available, with 41741 H\,{\footnotesize I} sources detected over a range of 7600 square degrees \citep{Zhang2024}. 
Since FASHI is observed in schedule-filler time, it is difficult to control a uniform distribution over the sky area, so we made additional observations of some regions of UMa with the on-the-fly (OTF) mode in January 2024 to ensure that every position in the UMa region has been observed at least twice.

FAST's 19-beam receiver \citep{Dunning2017} makes blind surveys more efficient and faster.
At 1.4 GHz, it has a half-power beamwidth (HPBW) of about 2.9\arcmin per beam.
The base system temperature in drift-scan mode is 16-19 K for different frequencies and increases with increasing ZA. 
The pointing error is less than 16\arcsec. 
The spectral-line backend is used with Spec(W), which provides two linear polarisation information for H\,{\footnotesize I} observations. 
It has 65536 channels in the 500 MHz bandwidth, resulting in frequency and velocity resolutions up to 7.6 kHz and 1.6 km~s$^{-1}$. 
The observation sampling time is 1 s. 
For intensity calibration, we used a strong noise diode injected every 32 s during the observation, with a noise diode temperature of about 11 K. 
The majority of the observing mode in the FASHI project is in drift-scan mode (DecDriftWithAngle). 
The MultiBeamOTF mode was also combined for directional observations of the UMa region.

\subsection{Data Processing}
\label{sec:dp} 

\begin{figure}
	\includegraphics[width=\columnwidth]{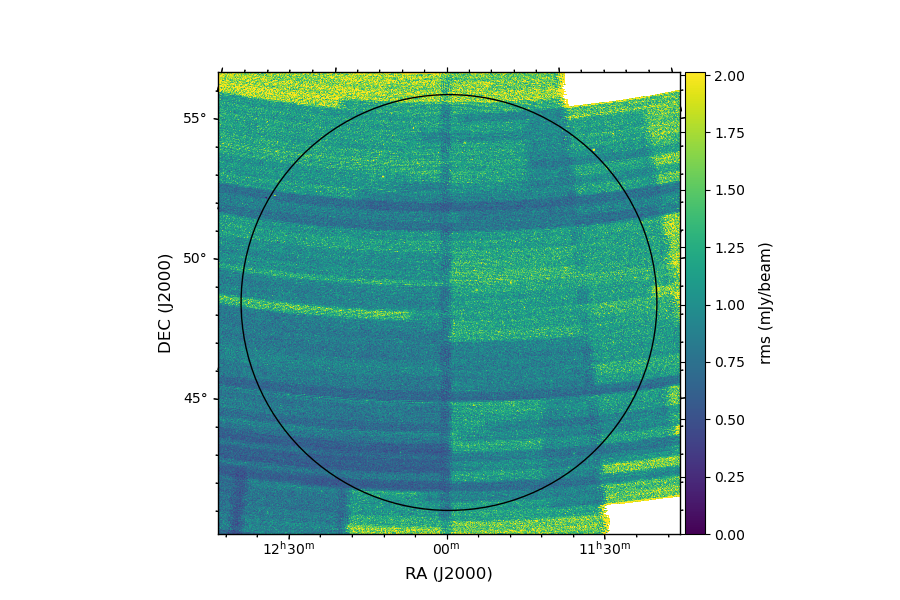}
    \caption{RMS distribution of FAST observations of the UMa region. 
    The black circle marks the UMa region with a radius of 7.5\degr.
    The mean value of RMS in the region is about 1 mJy~beam$^{-1}$.}
    \label{fig:rms}
\end{figure}

Data processing is mainly based on HiFAST pipeline \citep{Jing2024}, which is a dedicated pipeline for processing FAST H\,{\footnotesize I} data. 
Its processing steps are antenna temperature calibration, position determination, flux calibration, baseline removal, radio frequency interference (RFI) flagging, standing-wave removal, Doppler velocity calibration, and gridding. 
Each step is independently modular, and users can choose to use it according to their needs.
Firstly, the antenna temperature calibration is based on the noise diode temperature, which is frequency- and beam-dependent, and converts the machine counts to the antenna temperature.
Then the pointing position of each sampling point in the sky is calculated based on the position information of the feed. 
The flux calibration is obtained by calculating the antenna gain, which is approximately 16.1 K~Jy$^{-1}$, depending on ZA. 
For the removal of the baseline, we used the asymmetrically reweighted penalized least-squares (arPLS) algorithm \citep{Baek2015}, which iteratively increases the weight of the signal below the fitted baseline and decreases the weight above it to obtain the optimal fitted line. 
This method is very effective for studying H\,{\footnotesize I} emission lines and has performed well in many previous FAST works \citep{Xu2021, Yu2023, Zhou2023}.
We combined standing-wave removal with the sinusoidal function fitting method in the baseline removal module, which is effective for our data in the short frequency range between 1399-1429 MHz. 
There is no RFI interference in this range, so we did not use the RFI flagging module. 
After finally converting the frequencies to heliocentric velocities, we gridded them into a cube with a spatial resolution of 1\arcmin. 
In order to obtain high-quality spectral lines, we also performed a baseline removal and Hanning smoothing of the spectral lines within each pixel using the baseline removal module. 
The final velocity resolution of our spectra is about 4.9 km~s$^{-1}$.
Figure \ref{fig:rms} shows the distribution of the RMS within the final cube, which has a mean value of approximately 1 mJy~beam$^{-1}$. 
The inhomogeneity of the figure is mainly due to the difference in the number of observations, with a higher number of observations resulting in a lower RMS.

For the extraction of H\,{\footnotesize I} sources, we use the H\,{\footnotesize I} Source Finding Application (SoFiA) in version 2 \citep{Serra2015, Westmeier2021}. 
This is an application that finds and parameterizes H\,{\footnotesize I} sources in a 3D spectral line cube. 
The finding algorithm we used is the Smoothing + Clipping (S+C) algorithm \citep{Serra2012}. 
The algorithm involves data smoothing of a specified 3D kernel to pick out voxels whose absolute values exceed a certain threshold.
We set the threshold at 4$\sigma$, with $\sigma$ being the noise value around the detected source. 
After searching for sources, SoFiA also provides information on the location, spectra, and moment maps of these sources. 
Although SoFiA can quickly and automatically find H\,{\footnotesize I} sources in 3D data, there will be some fake sources, interferences, confusions, etc. in the results.
So manual screening afterward is unavoidable. 
We performed secondary processing on the sources filtered out by SoFiA, which includes removing the fake sources and interferences among them; fitting the H\,{\footnotesize I} sources with busy-functions \citep{Westmeier2014}, calculating the flux, line width, and velocity information of the spectral lines; and distinguishing the single H\,{\footnotesize I} sources in the confusions as much as possible.

\section{The Ursa Major Supergroup Catalog}
\label{sec:UMa catalog}

\subsection{Selection of Detections}
\label{sec:selection}

\begin{figure}
	\includegraphics[width=\columnwidth]{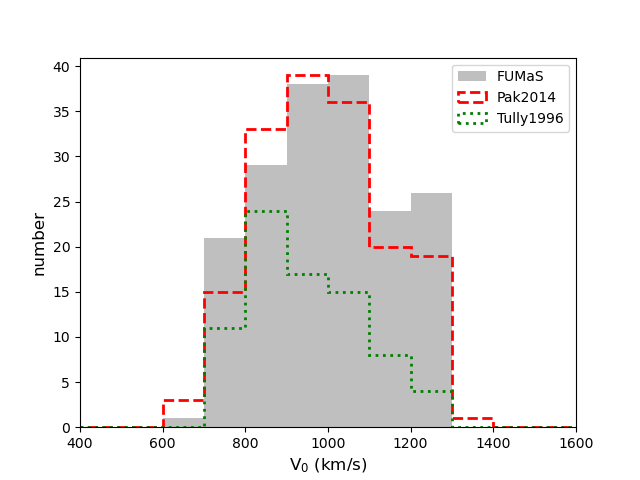}
    \caption{Local group velocity distribution of galaxies in the UMa supergroup. 
    The grey histogram is from FUMaS, the red dashed line is from \citet{Pak2014}, and the green dotted line is from \citet{Tully1996}.}
    \label{fig:V0}
\end{figure}

In the data processing described in the previous section, the positional range of the cubes we finished processing with HiFAST was extended outwards by a distance of about 8 degrees from the center of UMa.
In order to remove Galactic effects, we intercepted a velocity range of 100-2500 km~s$^{-1}$. 
Within this range, we used SoFiA combined with manual screening to make a preliminary selection of 318 detections. 
We then restricted the range to 7.5 degrees and selected members of the UMa supergroup using a biweighted scale estimator \citep{Beers1990}, a method used in the work of \citet{Pak2014}. 
By iteratively removing galaxies outside the mean velocity of 2$\sigma$ until convergence, we obtained a UMa supergroup catalog with 173 detections. 
Compared to the optical members, we identified 5 additional H\,{\footnotesize I} sources, bringing the total number of detections in our catalog to 178.
For comparison with previous data, we calculated the Local Group velocities of the detections via the equation: $V_\mathrm{0}=V_\mathrm{helio}+$\,300sin$l$\,cos$b$, where $l$ and $b$ are Galactic longitude and Galactic latitude, respectively.
The Local Group velocity range and dispersion of the detections in our catalog are 695.0-1291.3 km~s$^{-1}$ and $\sigma$ = 151.4 km~s$^{-1}$, respectively, as shown in Figure \ref{fig:V0}. 
This agrees well with the results of both \citet{Tully1996} and \citet{Pak2014}.

\subsection{The Ursa Major Supergroup Catalog}
\label{sec:sub UMa catalog}

Table \ref{tab:UMa catalog1} lists 146 H\,{\footnotesize I} sources with known-redshift optical counterparts of the UMa supergroup, including 140 individual galaxies, 3 confused pairs, and 3 confused small groups.
The remaining 32 detections without known-redshift optical counterparts are listed in Table \ref{tab:UMa catalog2}. 
Information on each parameter column in the table is described below:

Column 1: The index number of each UMa supergroup H\,{\footnotesize I} source.

Columns 2 and 3: Right ascension (RA) and declination (DEC) (J2000) of the detections. 
The uncertainty is calculated by the formula \citep{Wolfinger2013}: $\sigma_\mathrm{pos}=\mathrm{HPBW}/(F_\mathrm{peak}/\sigma_\mathrm{rms})$, where HPBW is the FAST beam size of 2.9\arcmin and $F_\mathrm{peak}$ with $\sigma_\mathrm{rms}$ see columns 7 and 11.

Column 4: Heliocentric velocity of the detections in km~s$^{-1}$.
The uncertainty is calculated as: $\sigma_\mathrm{v}=3\sqrt{P\delta v}/\mathrm{SNR}$, where $P$ is $(W_{20}-W_{50})/2$, $\delta v=4.9$ km~s$^{-1}$ is the velocity resolution and SNR is the signal-to-noise ratio as detailed in column 12 \citep{Koribalski2004}.

Column 5: Velocity widths of 50\% of the peak intensity of the spectral lines fitted by the busy-function.
Its uncertainty is calculated from $\sigma_{50}=2\sigma_\mathrm{v}$ \citep{Fouque1990}.

Column 6: Velocity widths of 20\% of the peak intensity of the spectral lines fitted by the busy-function. 
Its uncertainty is calculated from $\sigma_{20}=3\sigma_\mathrm{v}$.

Column 7: Peak intensity of the spectrum from the busy-function fit in mJy. 
Uncertainty is $\sigma_\mathrm{rms}$ in column 11.

Column 8: Integrated flux density of the busy-function fitted H\,{\footnotesize I} profile in Jy~km~s$^{-1}$. 
Subsequent SNR and mass are calculated based on this value. 

Column 9: Integrated flux density of the H\,{\footnotesize I} profile by summing the fluxes over the same velocity channels as $S_\mathrm{bf}$.

Column 10: Uncertainty in flux density. 
Calculated as $\sigma_\mathrm{s}=\sqrt{N}\delta v\sigma_\mathrm{rms}$, where $N$ is the number of velocity channels. 
Since the velocity channels of $S_\mathrm{bf}$ and $S_\mathrm{sum}$ are the same, the uncertainties are the same.

Column 11: Noise level of the H\,{\footnotesize I} spectrum in mJy at 4.9 km~s$^{-1}$ spectral resolution. 
The data used for the calculations are the signal-free spectral line regions.

Column 12: Signal-to-noise ratio of the detections, estimated by the following equation:
\begin{equation}
\label{eq:SNR}
\mathrm{SNR}=(\frac{S_\mathrm{bf}}{W_{50}})\frac{w^{1/2}_\mathrm{smo}}{\sigma_\mathrm{rms}}
\end{equation}
where $w_\mathrm{smo}=W_{50}/\delta v$ is a smoothing width \citep{Haynes2018}.

Column 13: Logarithm of the detection mass in units of solar mass. 
Calculated by the following formula:
\begin{equation}
M_{\mathrm{HI}}=2.356\times10^5D^2S_\mathrm{bf},
\label{eq:mass}
\end{equation}
where D is the distance, and due to the uniform distribution of the UMa galaxy supergroup, we adopt the average UMa distance of 17.4 Mpc \citep{Tully2012} for all the detections.

Columns 14-17: Information on the optical counterparts of the detections obtained from NED\footnote{\url{https://ned.ipac.caltech.edu/}} and SDSS DR16 \citep{Ahumada2020}, which are name, RA, DEC, and velocity.

Columns 18: Flags for newly detected H\,{\footnotesize I} sources (n), clouds on the outskirts of galaxies (c), sources in the first release of FASHI (f).

\setlength{\tabcolsep}{1.3mm}{
\begin{longrotatetable}
\begin{deluxetable*}{lccccccccccccccccc}
\label{tab:UMa catalog1}
\tabletypesize{\tiny}
\centerwidetable
\tablecaption{FUMaS H\,{\footnotesize I} source with known-redshift optical counterpart catalog.}
\tablehead{\colhead{[1]} & \colhead{[2]} & \colhead{[3]} & \colhead{[4]} & \colhead{[5]} & \colhead{[6]} & \colhead{[7]} & \colhead{[8]} & \colhead{[9]} & \colhead{[10]} & \colhead{[11]} & \colhead{[12]} & \colhead{[13]} & \colhead{[14]} & \colhead{[15]} & \colhead{[16]} & \colhead{[17]} & \colhead{[18]} \\ 
\colhead{ID} & \colhead{Ra} & \colhead{Dec} & \colhead{$V_\mathrm{helio}\pm\sigma_\mathrm{v}$} & \colhead{$W_{50}\pm\sigma_{50}$} & \colhead{$W_{20}\pm\sigma_{20}$} & \colhead{$F_\mathrm{peak}$} & \colhead{$S_\mathrm{bf}$} & \colhead{$S_\mathrm{sum}$} & \colhead{$\sigma_\mathrm{s}$} & \colhead{$\sigma_\mathrm{rms}$} & \colhead{SNR} & \colhead{log$M$} & \colhead{Opt name} & \colhead{Opt ra} & \colhead{Opt dec} & \colhead{Opt v} & \colhead{Flags} \\ 
\colhead{} & \colhead{(hms)} & \colhead{(dms)} & \colhead{(km~s$^{-1}$)} & \colhead{(km~s$^{-1}$)} & \colhead{(km~s$^{-1}$)} & \colhead{(mJy)} & \colhead{(Jy~km~s$^{-1}$)} & \colhead{(Jy~km~s$^{-1}$)} & \colhead{(Jy~km~s$^{-1}$)} & \colhead{(mJy)} & \colhead{} & \colhead{(M$_{\sun}$)} & \colhead{} & \colhead{(hms)} & \colhead{(dms)} & \colhead{(km~s$^{-1}$)} & \colhead{} \\ }
\startdata
FUMa 001 & 12:10:56.59 & 50:17:07.9 & $876.77\pm0.12$ & $102.18\pm0.24$ & $115.30\pm0.36$ & 61.3 & 6.267 & 6.334 & 0.090 & 2.1 & 136.2 & 8.7 & UGC 7176 & 12:10:55.10 & 50:17:16.1 & 888 & f\\
FUMa 002 & 12:11:05.33 & 50:29:08.1 & $772.33\pm0.03$ & $397.64\pm0.06$ & $424.35\pm0.09$ & 432.6 & 117.463 & 117.866 & 0.242 & 3.5 & 768.5 & 9.9 & NGC 4157 & 12:11:04.36 & 50:29:06.0 & 771 & f\\
FUMa 003 & 12:05:24.91 & 50:21:27.7 & $748.85\pm0.09$ & $278.01\pm0.18$ & $297.89\pm0.27$ & 85.6 & 23.182 & 23.371 & 0.162 & 2.7 & 235.9 & 9.2 & NGC 4085 & 12:05:22.74 & 50:21:10.3 & 760 & f\\
FUMa 004 & 12:05:38.77 & 50:32:31.3 & $755.17\pm0.03$ & $343.67\pm0.06$ & $373.13\pm0.09$ & 437.8 & 138.693 & 138.672 & 0.227 & 3.5 & 981.0 & 10.0 & NGC 4088 & 12:05:34.28 & 50:32:21.8 & 746 & f\\
FUMa 005 & 12:03:09.74 & 44:31:55.1 & $703.78\pm0.05$ & $235.46\pm0.10$ & $265.64\pm0.15$ & 189.3 & 41.872 & 41.764 & 0.150 & 2.6 & 469.0 & 9.5 & NGC 4051 & 12:03:09.61 & 44:31:52.7 & 700 & \\
FUMa 006 & 11:26:40.15 & 53:44:49.3 & $643.56\pm0.03$ & $131.91\pm0.06$ & $146.44\pm0.09$ & 418.0 & 44.822 & 44.987 & 0.140 & 3.0 & 595.4 & 9.5 & UGC 6446 & 11:26:40.46 & 53:44:49.0 & 646 & \\
FUMa 007 & 11:58:02.50 & 51:20:34.2 & $624.97\pm0.43$ & $25.53\pm0.86$ & $40.66\pm1.29$ & 32.2 & 0.961 & 0.860 & 0.069 & 2.0 & 42.9 & 7.8 & SDSS J115802.14+512057.7 & 11:58:02.14 & 51:20:57.8 & 576 & \\
FUMa 008 & 12:00:34.56 & 47:46:20.9 & $660.37\pm0.19$ & $81.59\pm0.38$ & $99.14\pm0.57$ & 42.0 & 3.430 & 3.509 & 0.070 & 1.7 & 103.6 & 8.4 & UGCA 262 & 12:00:35.43 & 47:46:25.5 & 569 & n\\
FUMa 009 & 12:25:34.40 & 45:41:01.0 & $714.86\pm0.14$ & $161.70\pm0.28$ & $186.42\pm0.42$ & 54.6 & 8.007 & 8.025 & 0.083 & 1.7 & 171.6 & 8.8 & NGC 4389 & 12:25:35.16 & 45:41:04.6 & 718 & f\\
FUMa 010 & 11:58:57.84 & 44:11:31.5 & $680.91\pm0.73$ & $68.25\pm1.46$ & $90.38\pm2.19$ & 13.1 & 0.732 & 0.781 & 0.053 & 1.3 & 30.1 & 7.7 & CGCG 215-13 & 11:58:56.87 & 44:11:33.4 & 687 & n\\
FUMa 011 & 11:58:30.33 & 43:56:33.0 & $833.00\pm0.06$ & $392.58\pm0.12$ & $412.93\pm0.18$ & 162.4 & 41.391 & 41.402 & 0.202 & 2.9 & 325.3 & 9.5 & NGC 4013 & 11:58:31.31 & 43:56:50.7 & 831 & \\
FUMa 012 & 12:16:43.48 & 46:04:43.9 & $704.40\pm0.11$ & $140.33\pm0.22$ & $153.01\pm0.33$ & 63.5 & 7.356 & 7.414 & 0.091 & 1.9 & 149.2 & 8.7 & UGC 7301 & 12:16:42.06 & 46:04:43.6 & 706 & f\\
FUMa 013 & 12:16:11.88 & 47:52:33.1 & $948.21\pm0.39$ & $338.23\pm0.78$ & $356.59\pm1.17$ & 19.6 & 3.346 & 3.506 & 0.104 & 1.6 & 51.7 & 8.4 & NGC 4220 & 12:16:11.73 & 47:53:00.4 & 914 & f\\
FUMa 014 & 12:15:46.45 & 48:07:46.0 & $725.32\pm0.10$ & $108.96\pm0.20$ & $154.27\pm0.30$ & 76.0 & 8.468 & 8.388 & 0.051 & 1.1 & 324.8 & 8.8 & NGC 4218 & 12:15:46.42 & 48:07:51.9 & 730 & f\\
FUMa 015 & 12:02:44.16 & 45:11:28.6 & $711.47\pm0.35$ & $158.53\pm0.70$ & $168.99\pm1.05$ & 11.8 & 1.437 & 1.502 & 0.059 & 1.2 & 43.5 & 8.0 & UGC 7022 & 12:02:43.71 & 45:11:27.5 & 709 & n\\
FUMa 016 & 11:55:05.40 & 44:06:06.5 & $647.49\pm1.24$ & $25.82\pm2.48$ & $34.71\pm3.72$ & 4.8 & 0.126 & 0.103 & 0.027 & 1.0 & 11.3 & 7.0 & SDSS J115506.03+440611.7 & 11:55:06.04 & 44:06:11.8 & 628 & n\\
FUMa 017 & 11:53:40.82 & 47:51:26.4 & $803.12\pm0.03$ & $259.06\pm0.06$ & $287.69\pm0.09$ & 202.0 & 44.316 & 44.490 & 0.093 & 1.6 & 791.3 & 9.5 & NGC 3949 & 11:53:41.79 & 47:51:31.4 & 794 & f\\
FUMa 018 & 11:41:18.42 & 46:23:54.6 & $753.15\pm0.16$ & $85.07\pm0.32$ & $112.52\pm0.48$ & 56.2 & 4.803 & 4.814 & 0.065 & 1.5 & 152.4 & 8.5 & CGCG 242-075 & 11:41:22.01 & 46:23:35.5 & 746 & \\
FUMa 019 & 11:39:19.23 & 46:30:41.2 & $734.74\pm0.04$ & $90.69\pm0.08$ & $132.34\pm0.12$ & 440.3 & 41.399 & 40.951 & 0.120 & 2.8 & 697.5 & 9.5 & NGC 3782 & 11:39:20.75 & 46:30:49.2 & 733 & \\
FUMa 020 & 11:39:46.78 & 46:37:03.9 & $710.98\pm0.20$ & $42.11\pm0.40$ & $68.92\pm0.60$ & 26.8 & 1.272 & 1.244 & 0.031 & 0.7 & 119.9 & 8.0 & SDSS J113948.69+463711.4 & 11:39:48.83 & 46:37:11.6 & 712 & \\
FUMa 021 & 11:59:37.02 & 42:57:10.0 & $679.20\pm0.56$ & $55.12\pm1.12$ & $74.86\pm1.68$ & 10.4 & 0.580 & 0.576 & 0.036 & 0.9 & 37.3 & 7.6 & SDSS J115936.94+425715.7 & 11:59:36.95 & 42:57:15.8 & 668 & n\\
FUMa 022 & 11:48:35.81 & 43:43:26.8 & $739.20\pm0.28$ & $135.76\pm0.56$ & $150.14\pm0.84$ & 25.9 & 2.889 & 2.875 & 0.086 & 1.8 & 62.3 & 8.3 & UGC 6776 & 11:48:35.76 & 43:43:17.4 & 738 & \\
FUMa 023 & 11:45:54.44 & 50:12:02.4 & $753.02\pm0.14$ & $96.19\pm0.28$ & $135.17\pm0.42$ & 88.8 & 8.722 & 8.716 & 0.080 & 1.9 & 215.7 & 8.8 & NGC 3870 & 11:45:56.62 & 50:11:59.4 & 754 & \\
FUMa 024 & 12:17:32.49 & 47:59:37.7 & $694.58\pm0.29$ & $37.62\pm0.58$ & $57.31\pm0.87$ & 23.0 & 0.916 & 0.966 & 0.034 & 0.9 & 72.1 & 7.8 & 2MASX J12173195+4759420 & 12:17:31.88 & 47:59:43.1 & 687 & f\\
FUMa 025 & 12:06:27.04 & 52:42:07.1 & $846.42\pm0.20$ & $303.38\pm0.40$ & $344.11\pm0.60$ & 37.8 & 11.470 & 11.432 & 0.124 & 2.0 & 150.8 & 8.9 & NGC 4102 & 12:06:23.05 & 52:42:39.7 & 785 & f\\
FUMa 026 & 11:59:51.29 & 55:39:52.4 & $736.48\pm0.44$ & $115.95\pm0.88$ & $129.04\pm1.32$ & 15.1 & 1.751 & 1.761 & 0.087 & 1.9 & 38.7 & 8.1 & UGC 6988 & 11:59:51.71 & 55:39:55.2 & 732 & \\
FUMa 027 & 11:57:51.51 & 55:26:55.7 & $1000.60\pm1.07$ & $583.63\pm2.14$ & $692.08\pm3.21$ & 22.2 & 8.223 & 8.342 & 0.227 & 3.4 & 45.6 & 8.8 & NGC 3998 & 11:57:56.13 & 55:27:12.9 & 1020 & \\
FUMa 028 & 11:55:45.37 & 55:19:13.8 & $835.14\pm0.08$ & $253.29\pm0.16$ & $278.50\pm0.24$ & 111.4 & 17.855 & 17.923 & 0.099 & 1.7 & 298.8 & 9.1 & NGC 3972 & 11:55:45.13 & 55:19:14.1 & 842 & \\
FUMa 029 & 11:56:28.65 & 55:07:07.9 & $1112.19\pm0.06$ & $212.16\pm0.12$ & $245.07\pm0.18$ & 151.0 & 28.601 & 28.919 & 0.114 & 2.1 & 425.7 & 9.3 & NGC 3982 & 11:56:28.13 & 55:07:30.8 & 1122 & f\\
FUMa 030 & 12:00:19.05 & 50:39:12.2 & $743.78\pm0.42$ & $126.78\pm0.84$ & $140.17\pm1.26$ & 13.9 & 1.413 & 1.415 & 0.064 & 1.4 & 41.2 & 8.0 & UGC 6992 & 12:00:18.89 & 50:39:10.6 & 759 & \\
FUMa 031 & 11:57:18.59 & 49:16:48.1 & $777.33\pm0.02$ & $121.55\pm0.04$ & $137.60\pm0.06$ & 393.2 & 42.575 & 42.759 & 0.107 & 2.3 & 759.3 & 9.5 & UGC 6930 & 11:57:17.3 & 49:16:59.5 & 777 & \\
FUMa 032 & 11:33:21.78 & 47:01:43.9 & $861.97\pm0.04$ & $269.72\pm0.08$ & $290.92\pm0.12$ & 545.3 & 111.991 & 112.843 & 0.387 & 6.4 & 479.9 & 9.9 & NGC 3726 & 11:33:21.11 & 47:01:45.2 & 864 & f\\
FUMa 033 & 12:20:16.72 & 45:54:06.4 & $730.88\pm0.13$ & $67.18\pm0.26$ & $80.05\pm0.39$ & 43.2 & 2.719 & 2.779 & 0.048 & 1.2 & 125.0 & 8.3 & UGC 7391 & 12:20:16.24 & 45:54:30.2 & 737 & f\\
FUMa 034 & 11:23:21.66 & 50:53:26.4 & $791.12\pm0.10$ & $173.88\pm0.20$ & $186.67\pm0.30$ & 83.3 & 11.941 & 11.956 & 0.129 & 2.5 & 164.1 & 8.9 & UGC 6399 & 11:23:23.22 & 50:53:33.8 & 791 & \\
FUMa 035 & 12:11:34.81 & 47:39:17.2 & $742.53\pm0.44$ & $80.10\pm0.88$ & $93.34\pm1.32$ & 14.3 & 1.087 & 1.085 & 0.060 & 1.4 & 38.5 & 7.9 & SDSS J121134.99+473927.1 & 12:11:35.00 & 47:39:27.2 & 752 & f\\
FUMa 036 & 12:20:47.52 & 47:49:33.8 & $742.65\pm0.22$ & $52.69\pm0.44$ & $68.99\pm0.66$ & 48.8 & 2.297 & 2.290 & 0.065 & 1.7 & 82.7 & 8.2 & UGC 7401 & 12:20:48.38 & 47:49:33.6 & 762 & f\\
FUMa 037 & 11:55:14.90 & 44:09:02.2 & $726.44\pm0.32$ & $41.36\pm0.64$ & $52.52\pm0.96$ & 32.8 & 1.358 & 1.427 & 0.072 & 2.0 & 48.2 & 8.0 & KDG 081 & 11:55:13.99 & 44:09:00.4 & 727 & \\
FUMa 038 & 11:46:07.66 & 47:29:30.1 & $887.42\pm0.08$ & $334.52\pm0.16$ & $356.14\pm0.24$ & 124.8 & 25.762 & 25.854 & 0.153 & 2.3 & 272.1 & 9.3 & NGC 3877 & 11:46:07.73 & 47:29:40.4 & 895 & \\
FUMa 039 & 12:09:29.12 & 43:41:04.5 & $887.97\pm0.24$ & $261.26\pm0.48$ & $341.82\pm0.72$ & 101.6 & 26.648 & 26.391 & 0.254 & 4.3 & 174.2 & 9.3 & NGC 4138 & 12:09:29.80 & 43:41:07.1 & 874 & f\\
FUMa 040 & 12:12:55.71 & 52:15:50.8 & $781.81\pm0.17$ & $99.52\pm0.34$ & $120.73\pm0.51$ & 48.6 & 4.753 & 4.779 & 0.075 & 1.7 & 126.4 & 8.5 & UGC 7218 & 12:12:56.88 & 52:15:54.4 & 770 & f\\
FUMa 041 & 11:50:46.31 & 45:48:23.4 & $805.01\pm0.09$ & $154.23\pm0.18$ & $173.66\pm0.27$ & 114.7 & 16.161 & 16.190 & 0.132 & 2.7 & 222.2 & 9.1 & UGC 6818 & 11:50:46.95 & 45:48:25.9 & 808 & \\
FUMa 042 & 11:52:50.30 & 44:07:13.9 & $808.18\pm0.02$ & $92.68\pm0.04$ & $109.61\pm0.06$ & 1097.8 & 94.685 & 95.660 & 0.173 & 4.0 & 1109.6 & 9.8 & NGC 3938 & 11:52:49.43 & 44:07:14.8 & 808 & f\\
FUMa 043 & 11:33:48.90 & 53:07:15.6 & $1045.30\pm0.19$ & $230.14\pm0.38$ & $249.79\pm0.57$ & 37.3 & 5.451 & 5.513 & 0.086 & 1.5 & 107.6 & 8.6 & NGC 3729 & 11:33:49.38 & 53:07:32.0 & 1000 & \\
FUMa 044 & 11:32:36.96 & 53:04:19.4 & $987.10\pm0.03$ & $459.62\pm0.06$ & $478.50\pm0.09$ & 658.6 & 176.544 & 177.016 & 0.341 & 4.6 & 808.8 & 10.1 & NGC 3718 & 11:32:34.85 & 53:04:04.5 & 993 & f\\
FUMa 045 & 12:20:19.10 & 48:08:21.2 & $797.25\pm0.13$ & $125.46\pm0.26$ & $139.86\pm0.39$ & 35.2 & 3.957 & 3.966 & 0.054 & 1.1 & 139.5 & 8.5 & UGC 7392 & 12:20:17.48 & 48:08:14.9 & 808 & f\\
FUMa 046 & 11:56:48.90 & 50:48:54.2 & $890.02\pm0.09$ & $131.18\pm0.18$ & $141.84\pm0.27$ & 106.1 & 10.939 & 11.117 & 0.126 & 2.7 & 162.8 & 8.9 & UGC 6922 & 11:56:52.19 & 50:49:01.4 & 877 & \\
FUMa 047 & 11:59:52.56 & 50:29:55.0 & $939.28\pm0.27$ & $81.09\pm0.54$ & $92.79\pm0.81$ & 17.4 & 1.317 & 1.336 & 0.046 & 1.1 & 59.5 & 8.0 & SDSS J115950.81+502955.3 & 11:59:50.82 & 50:29:55.3 & 876 & \\
FUMa 048 & 11:39:47.52 & 54:30:45.4 & $795.11\pm0.59$ & $108.46\pm1.18$ & $121.03\pm1.77$ & 13.8 & 1.101 & 1.141 & 0.076 & 1.7 & 28.2 & 7.9 & SDSS J113948.41+543116.0 & 11:39:48.48 & 54:31:15.1 & 790 & n\\
FUMa 049 & 11:58:38.20 & 47:15:34.0 & $901.52\pm0.04$ & $263.86\pm0.08$ & $283.31\pm0.12$ & 182.5 & 42.798 & 42.908 & 0.153 & 2.6 & 463.7 & 9.5 & NGC 4010 & 11:58:37.97 & 47:15:41.2 & 902 & f\\
FUMa 050 & 11:55:25.51 & 54:39:26.0 & $847.09\pm0.26$ & $127.01\pm0.52$ & $145.13\pm0.78$ & 41.1 & 4.971 & 4.927 & 0.120 & 2.6 & 77.6 & 8.5 & UGC 6894 & 11:55:23.50 & 54:39:26.5 & 850 & \\
FUMa 051 & 11:48:45.10 & 49:21:30.8 & $800.71\pm0.70$ & $37.74\pm1.40$ & $58.08\pm2.10$ & 21.3 & 0.863 & 0.889 & 0.075 & 2.1 & 30.3 & 7.8 & SDSS J114845.15+492129.6 & 11:48:45.17 & 49:21:29.3 & 798 & n\\
FUMa 052 & 11:58:12.21 & 48:52:55.3 & $820.54\pm0.26$ & $79.36\pm0.52$ & $92.70\pm0.78$ & 31.5 & 2.501 & 2.522 & 0.078 & 1.9 & 66.4 & 8.3 & MCG +08-22-048 & 11:58:11.46 & 48:52:53.7 & 833 & \\
FUMa 053 & 11:49:37.82 & 48:24:58.0 & $961.86\pm0.11$ & $35.71\pm0.22$ & $52.76\pm0.33$ & 162.1 & 6.045 & 6.011 & 0.097 & 2.7 & 169.5 & 8.6 & NGC 3906 & 11:49:40.67 & 48:25:33.9 & 959 & \\
FUMa 054 & 11:49:33.29 & 44:23:51.0 & $795.42\pm0.78$ & $32.57\pm1.56$ & $47.17\pm2.34$ & 12.9 & 0.433 & 0.446 & 0.053 & 1.5 & 22.9 & 7.5 & SDSS J114930.92+442433.1 & 11:49:30.95 & 44:24:33.6 & 802 & n\\
FUMa 055 & 11:50:49.97 & 48:14:57.5 & $807.79\pm1.45$ & $52.57\pm2.90$ & $69.32\pm4.35$ & 4.8 & 0.230 & 0.200 & 0.041 & 1.1 & 13.2 & 7.2 & Mrk 1460 & 11:50:50.05 & 48:15:04.9 & 810 & \\
FUMa 056 & 11:59:45.14 & 53:37:13.4 & $1138.91\pm0.64$ & $27.76\pm1.28$ & $47.97\pm1.92$ & 14.2 & 0.438 & 0.475 & 0.039 & 1.1 & 33.0 & 7.5 & SDSS J115943.26+533638.9 & 11:59:43.27 & 53:36:39.0 & 1132 & \\
FUMa 057 & 11:40:37.71 & 46:07:27.8 & $885.94\pm0.78$ & $41.81\pm1.56$ & $70.95\pm2.34$ & 11.3 & 0.537 & 0.542 & 0.043 & 1.2 & 32.3 & 7.6 & SDSS J114035.57+460727.5 & 11:40:35.56 & 46:07:27.6 & 888 & \\
FUMa 058 & 11:40:07.04 & 45:56:09.9 & $848.97\pm0.02$ & $39.62\pm0.04$ & $54.42\pm0.06$ & 693.8 & 27.682 & 28.023 & 0.080 & 2.2 & 913.3 & 9.3 & UGC 6628 & 11:40:05.78 & 45:56:34.1 & 849 & \\
FUMa 059 & 12:06:39.62 & 54:45:48.6 & $846.86\pm0.16$ & $86.24\pm0.32$ & $96.88\pm0.48$ & 33.1 & 2.855 & 2.884 & 0.060 & 1.5 & 95.6 & 8.3 & SDSS J120637.89+544557.7 & 12:06:37.92 & 54:45:57.9 & 850 & f\\
FUMa 060 & 11:56:29.70 & 50:25:34.0 & $914.10\pm0.05$ & $188.04\pm0.10$ & $202.82\pm0.15$ & 212.6 & 30.565 & 30.732 & 0.152 & 2.9 & 353.1 & 9.3 & UGC 6917 & 11:56:26.43 & 50:25:43.6 & 911 & \\
FUMa 061 & 11:58:50.12 & 46:27:51.4 & $819.15\pm0.35$ & $30.33\pm0.70$ & $47.72\pm1.05$ & 24.6 & 0.817 & 0.815 & 0.042 & 1.2 & 56.2 & 7.8 & SDSS J115849.68+462753.0 & 11:58:49.68 & 46:27:53.1 & 792 & n\\
FUMa 062 & 12:13:16.65 & 43:41:58.3 & $931.74\pm0.02$ & $233.98\pm0.04$ & $246.62\pm0.06$ & 310.3 & 48.953 & 49.195 & 0.103 & 1.8 & 799.6 & 9.5 & NGC 4183 & 12:13:16.88 & 43:41:54.9 & 929 & f\\
FUMa 063 & 11:50:45.92 & 51:49:27.2 & $955.67\pm0.04$ & $278.15\pm0.08$ & $291.75\pm0.12$ & 159.7 & 28.860 & 28.901 & 0.103 & 1.9 & 406.9 & 9.3 & NGC 3917 & 11:50:45.49 & 51:49:28.6 & 965 & \\
FUMa 064 & 11:49:54.60 & 51:44:21.3 & $946.16\pm0.75$ & $41.78\pm1.50$ & $60.47\pm2.25$ & 7.0 & 0.300 & 0.297 & 0.047 & 0.8 & 27.1 & 7.3 & SBS 1147+520 & 11:49:54.46 & 51:44:11.3 & 938 & \\
FUMa 065 & 11:50:48.70 & 56:27:32.9 & $881.79\pm0.10$ & $117.23\pm0.20$ & $132.92\pm0.30$ & 161.1 & 14.846 & 14.866 & 0.158 & 3.5 & 179.5 & 9.0 & UGC 6816 & 11:50:47.53 & 56:27:19.8 & 894 & \\
FUMa 066 & 11:16:58.78 & 50:35:19.2 & $847.55\pm0.33$ & $57.06\pm0.66$ & $75.40\pm0.99$ & 23.1 & 1.327 & 1.247 & 0.038 & 1.3 & 60.6 & 8.0 & CGCG 268-012 & 11:17:00.29 & 50:35:05.2 & 860 & n\\
FUMa 067 & 12:05:48.30 & 50:47:38.1 & $827.22\pm1.27$ & $28.50\pm2.54$ & $43.32\pm3.81$ & 4.8 & 0.146 & 0.154 & 0.029 & 0.9 & 14.2 & 7.0 & SDSS J120549.54+504729.0 & 12:05:49.55 & 50:47:29.1 & 827 & n\\
FUMa 068 & 11:53:49.29 & 52:19:51.0 & $1048.75\pm0.04$ & $407.50\pm0.08$ & $424.03\pm0.12$ & 201.8 & 52.217 & 52.344 & 0.168 & 2.4 & 491.7 & 9.6 & NGC 3953 & 11:53:49.03 & 52:19:36.5 & 1050 & f\\
FUMa 069 & 11:52:09.71 & 52:06:35.3 & $1017.01\pm0.04$ & $140.43\pm0.08$ & $153.93\pm0.12$ & 134.3 & 16.378 & 16.385 & 0.069 & 1.4 & 435.7 & 9.1 & UGC 6840 & 11:52:06.98 & 52:06:29.1 & 1015 & \\
FUMa 070 & 12:15:52.03 & 47:05:28.4 & $1029.17\pm0.05$ & $387.60\pm0.10$ & $404.69\pm0.15$ & 120.9 & 32.363 & 32.397 & 0.125 & 1.8 & 411.4 & 9.4 & NGC 4217 & 12:15:50.90 & 47:05:30.4 & 1028 & f\\
FUMa 071 & 11:53:56.86 & 50:10:24.6 & $854.76\pm1.14$ & $45.10\pm2.28$ & $54.86\pm3.42$ & 6.3 & 0.255 & 0.190 & 0.049 & 1.3 & 12.9 & 7.3 & UGC 6874 & 11:53:57.79 & 50:10:48.9 & 869 & n\\
FUMa 072 & 11:44:25.14 & 48:50:02.4 & $899.17\pm0.06$ & $91.21\pm0.12$ & $104.08\pm0.18$ & 205.6 & 16.852 & 17.021 & 0.118 & 2.7 & 292.5 & 9.1 & UGC 6713 & 11:44:24.92 & 48:50:07.8 & 899 & \\
FUMa 073 & 11:53:33.93 & 45:54:20.0 & $860.05\pm0.55$ & $25.34\pm1.10$ & $49.02\pm1.65$ & 26.7 & 0.850 & 0.858 & 0.064 & 1.8 & 41.4 & 7.8 & SDSS J115332.97+455421.8 & 11:53:32.98 & 45:54:21.9 & 802 & n\\
FUMa 074 & 11:56:43.14 & 48:20:04.7 & $952.19\pm0.08$ & $127.38\pm0.16$ & $172.19\pm0.24$ & 160.3 & 20.605 & 20.448 & 0.104 & 2.2 & 372.7 & 9.2 & NGC 3985 & 11:56:42.08 & 48:20:02.1 & 983 & \\
FUMa 075 & 11:59:28.40 & 42:21:34.2 & $888.30\pm0.30$ & $50.25\pm0.60$ & $77.33\pm0.90$ & 20.7 & 1.116 & 1.104 & 0.034 & 0.9 & 81.1 & 7.9 & KUG 1156+426 & 11:59:29.18 & 42:20:56.6 & 885 & n\\
FUMa 076 & 11:27:03.55 & 51:04:21.5 & $872.44\pm0.78$ & $34.60\pm1.56$ & $45.25\pm2.34$ & 9.3 & 0.324 & 0.326 & 0.045 & 1.3 & 19.6 & 7.4 & SDSS J112705.35+510428.9 & 11:27:05.40 & 51:04:28.8 & 886 & n\\
FUMa 077 & 12:23:07.79 & 53:00:54.6 & $890.65\pm0.44$ & $43.05\pm0.88$ & $63.84\pm1.32$ & 14.6 & 0.651 & 0.640 & 0.035 & 0.9 & 48.1 & 7.7 & SDSS J122307.98+530120.6 & 12:23:08.07 & 53:01:20.5 & 894 & f\\
FUMa 078 & 12:06:08.69 & 49:34:53.1 & $1077.26\pm0.05$ & $371.57\pm0.10$ & $395.25\pm0.15$ & 227.6 & 48.955 & 49.021 & 0.170 & 2.5 & 461.1 & 9.5 & NGC 4100 & 12:06:08.53 & 49:34:57.0 & 1074 & f\\
FUMa 079 & 12:15:37.42 & 44:16:56.7 & $894.97\pm0.38$ & $59.03\pm0.76$ & $74.17\pm1.14$ & 19.4 & 1.143 & 1.153 & 0.055 & 1.4 & 47.8 & 7.9 & SDSS J121537.12+441710.0 & 12:15:37.12 & 44:17:10.2 & 892 & f\\
FUMa 080 & 11:48:01.76 & 49:48:15.5 & $923.18\pm0.17$ & $95.95\pm0.34$ & $107.11\pm0.51$ & 71.5 & 6.651 & 6.627 & 0.140 & 3.3 & 94.9 & 8.7 & UGC 6773 & 11:48:00.26 & 49:48:30.5 & 924 & \\
FUMa 081 & 11:45:26.90 & 48:29:18.1 & $883.46\pm0.92$ & $36.67\pm1.84$ & $51.63\pm2.76$ & 9.0 & 0.336 & 0.358 & 0.045 & 1.3 & 19.7 & 7.4 & SDSS J114525.71+482906.8 & 11:45:25.66 & 48:29:07.7 & 892 & n\\
FUMa 082 & 12:06:25.06 & 42:26:07.0 & $878.99\pm0.84$ & $44.90\pm1.68$ & $58.17\pm2.52$ & 5.6 & 0.254 & 0.256 & 0.030 & 0.8 & 20.3 & 7.3 & SDSS J120625.35+422604.7 & 12:06:25.40 & 42:26:04.8 & 886 & f\\
FUMa 083 & 11:55:39.28 & 56:14:57.6 & $951.14\pm0.68$ & $123.59\pm1.36$ & $160.90\pm2.04$ & 16.9 & 2.095 & 2.034 & 0.096 & 2.0 & 41.8 & 8.2 & SBS 1153+565 & 11:55:37.18 & 56:15:11.9 & 948 & \\
FUMa 084 & 12:14:04.23 & 53:44:49.4 & $898.87\pm0.44$ & $36.28\pm0.88$ & $55.88\pm1.32$ & 24.8 & 0.968 & 0.958 & 0.055 & 1.5 & 47.2 & 7.8 & SBS 1211+540 & 12:14:02.49 & 53:45:17.5 & 903 & f\\
FUMa 085 & 11:56:03.00 & 52:26:38.5 & $879.67\pm0.70$ & $33.55\pm1.40$ & $41.80\pm2.10$ & 6.6 & 0.221 & 0.219 & 0.031 & 0.9 & 19.3 & 7.2 & SDSS J115603.70+522618.3 & 11:56:03.71 & 52:26:18.4 & 875 & n\\
FUMa 086 & 11:42:26.19 & 51:35:38.6 & $975.53\pm0.09$ & $181.28\pm0.18$ & $193.40\pm0.27$ & 90.2 & 11.639 & 11.697 & 0.108 & 2.1 & 189.3 & 8.9 & UGC 6667 & 11:42:26.30 & 51:35:53.2 & 973 & \\
FUMa 087 & 11:37:44.16 & 54:02:29.4 & $893.81\pm1.33$ & $48.67\pm2.66$ & $75.49\pm3.99$ & 4.4 & 0.228 & 0.209 & 0.029 & 0.8 & 18.3 & 7.2 & 2MASX J11374447+5402441 & 11:37:44.45 & 54:02:44.6 & 896 & \\
FUMa 088 & 11:48:18.28 & 56:19:40.2 & $1081.12\pm0.45$ & $19.92\pm0.90$ & $30.44\pm1.35$ & 31.6 & 0.669 & 0.635 & 0.068 & 2.0 & 33.8 & 7.7 & SDSS J114820.16+562045.7 & 11:48:20.16 & 56:20:45.5 & 1030 & n\\
FUMa 089 & 11:49:12.80 & 56:05:41.0 & $1158.52\pm0.12$ & $468.73\pm0.24$ & $488.61\pm0.36$ & 174.1 & 45.753 & 46.011 & 0.418 & 5.6 & 170.8 & 9.5 & NGC 3898 & 11:49:15.27 & 56:05:04.2 & 1162 & f\\
FUMa 090 & 11:49:30.89 & 56:01:39.8 & $921.21\pm1.25$ & $24.00\pm2.50$ & $31.83\pm3.75$ & 3.3 & 0.077 & 0.074 & 0.020 & 0.7 & 10.5 & 6.7 & SDSS J114929.60+560154.1 & 11:49:29.65 & 56:01:54.1 & 904 & n\\
FUMa 091 & 12:11:00.27 & 52:50:03.2 & $914.44\pm1.05$ & $47.00\pm2.10$ & $75.57\pm3.15$ & 8.6 & 0.443 & 0.431 & 0.046 & 1.2 & 23.9 & 7.5 & SBS 1208+531 & 12:11:00.70 & 52:49:57.4 & 911 & f\\
FUMa 092 & 11:43:32.36 & 55:28:48.9 & $967.61\pm0.21$ & $145.99\pm0.42$ & $155.59\pm0.63$ & 39.2 & 4.920 & 4.972 & 0.129 & 2.6 & 70.0 & 8.5 & UGC 6685 & 11:43:31.11 & 55:28:42.7 & 974 & \\
FUMa 093 & 12:01:00.32 & 49:54:44.4 & $936.38\pm0.22$ & $66.39\pm0.44$ & $76.25\pm0.66$ & 30.9 & 1.832 & 1.848 & 0.060 & 1.5 & 67.9 & 8.1 & UGC 6999 & 12:01:01.23 & 49:54:46.1 & 912 & \\
FUMa 094 & 11:50:39.71 & 55:21:32.8 & $950.79\pm0.06$ & $42.27\pm0.12$ & $57.29\pm0.18$ & 315.8 & 13.706 & 13.947 & 0.110 & 3.0 & 316.2 & 9.0 & NGC 3913 & 11:50:38.92 & 55:21:13.9 & 955 & \\
FUMa 095 & 11:46:26.35 & 53:24:28.8 & $917.24\pm1.83$ & $28.59\pm3.66$ & $36.34\pm5.49$ & 3.1 & 0.087 & 0.072 & 0.034 & 1.0 & 7.1 & 6.8 & SDSS J114628.27+532443.4 & 11:46:28.29 & 53:24:43.4 & 926 & n\\
FUMa 096 & 11:51:49.41 & 48:41:23.9 & $983.74\pm0.21$ & $67.56\pm0.42$ & $96.45\pm0.63$ & 138.5 & 9.275 & 9.162 & 0.170 & 4.2 & 121.3 & 8.8 & NGC 3928 & 11:51:47.63 & 48:40:59.7 & 978 & \\
FUMa 097 & 11:48:54.54 & 47:34:50.4 & $958.00\pm0.48$ & $29.46\pm0.96$ & $48.68\pm1.44$ & 27.6 & 0.907 & 0.818 & 0.063 & 1.8 & 43.0 & 7.8 & SDSS J114855.43+473457.5 & 11:48:55.45 & 47:34:57.5 & 963 & \\
FUMa 098 & 12:11:56.28 & 46:58:35.2 & $974.03\pm0.46$ & $97.97\pm0.92$ & $113.25\pm1.38$ & 11.0 & 0.894 & 0.891 & 0.045 & 1.0 & 40.0 & 7.8 & MCG +08-22-083 & 12:11:55.64 & 46:58:55.1 & 985 & f\\
FUMa 099 & 12:19:17.75 & 44:47:57.7 & $936.13\pm1.10$ & $41.94\pm2.20$ & $56.75\pm3.30$ & 3.5 & 0.149 & 0.141 & 0.022 & 0.6 & 16.4 & 7.0 & SDSS J121915.07+444801.8 & 12:19:15.11 & 44:48:02.7 & 922 & f\\
FUMa 100 & 11:51:00.79 & 47:56:31.2 & $945.64\pm1.12$ & $32.42\pm2.24$ & $37.67\pm3.36$ & 5.2 & 0.134 & 0.138 & 0.038 & 1.1 & 9.6 & 7.0 & SDSS J115059.60+475749.4 & 11:50:59.68 & 47:57:48.6 & 942 & \\
FUMa 101 & 12:10:46.85 & 45:43:19.8 & $943.19\pm0.65$ & $14.32\pm1.30$ & $33.00\pm1.95$ & 11.5 & 0.217 & 0.217 & 0.028 & 0.8 & 31.1 & 7.2 & LEDA 2832109 & 12:10:46.62 & 45:43:26.1 & 959 & f\\
FUMa 102 & 12:09:28.63 & 54:56:09.4 & $967.65\pm0.59$ & $39.83\pm1.18$ & $58.85\pm1.77$ & 19.6 & 0.814 & 0.818 & 0.061 & 1.7 & 34.9 & 7.8 & SDSS J120931.73+545618.1 & 12:09:31.81 & 54:56:19.0 & 995 & f\\
FUMa 103 & 11:52:38.74 & 50:02:19.3 & $999.91\pm0.16$ & $76.51\pm0.32$ & $99.05\pm0.48$ & 95.4 & 6.491 & 6.419 & 0.100 & 2.4 & 139.6 & 8.7 & UGC 6849 & 11:52:39.14 & 50:02:16.2 & 1003 & \\
FUMa 104 & 11:56:46.34 & 49:01:32.6 & $980.95\pm0.52$ & $48.39\pm1.04$ & $69.36\pm1.56$ & 16.4 & 0.817 & 0.827 & 0.048 & 1.3 & 41.3 & 7.8 & SDSS J115644.29+490118.2 & 11:56:44.34 & 49:01:17.8 & 984 & n\\
FUMa 105 & 12:12:56.18 & 44:05:23.9 & $957.54\pm1.23$ & $35.94\pm2.46$ & $50.04\pm3.69$ & 5.0 & 0.182 & 0.155 & 0.033 & 1.0 & 14.3 & 7.1 & SDSS J121255.18+440527.3 & 12:12:55.18 & 44:05:27.7 & 969 & f\\
FUMa 106 & 11:59:09.11 & 52:42:05.6 & $1083.07\pm0.02$ & $173.75\pm0.04$ & $189.02\pm0.06$ & 332.7 & 44.072 & 44.208 & 0.088 & 1.7 & 892.5 & 9.5 & UGC 6983 & 11:59:09.82 & 52:42:25.9 & 1081 & f\\
FUMa 107 & 11:46:12.73 & 54:10:07.8 & $983.99\pm0.70$ & $30.55\pm1.40$ & $33.87\pm2.10$ & 3.6 & 0.100 & 0.100 & 0.022 & 0.7 & 12.2 & 6.9 & SDSS J114613.44+541034.3 & 11:46:13.46 & 54:10:34.2 & 996 & \\
FUMa 108 & 11:55:49.27 & 45:10:03.4 & $1037.65\pm0.32$ & $77.14\pm0.64$ & $106.89\pm0.96$ & 44.5 & 3.483 & 3.436 & 0.093 & 2.3 & 79.7 & 8.4 & SDSS J115551.80+450945.5 & 11:55:51.81 & 45:09:45.6 & 1048 & n\\
FUMa 109 & 11:54:42.24 & 46:36:14.7 & $1004.18\pm0.61$ & $32.72\pm1.22$ & $46.54\pm1.83$ & 10.4 & 0.347 & 0.351 & 0.032 & 1.0 & 28.7 & 7.4 & SDSS J115441.22+463636.2 & 11:54:41.22 & 46:36:36.3 & 1009 & \\
FUMa 110 & 12:03:22.84 & 43:44:27.5 & $1042.54\pm0.41$ & $71.24\pm0.82$ & $87.53\pm1.23$ & 12.9 & 0.919 & 0.979 & 0.043 & 1.1 & 46.0 & 7.8 & 2MASX J12032304+4344395 & 12:03:23.01 & 43:44:40.0 & 1063 & \\
FUMa 111 & 12:09:48.35 & 43:14:02.4 & $1061.82\pm0.10$ & $71.13\pm0.20$ & $83.90\pm0.30$ & 52.2 & 3.590 & 3.781 & 0.045 & 1.1 & 173.8 & 8.4 & UGC 7146 & 12:09:49.15 & 43:14:04.9 & 1065 & f\\
FUMa 112 & 11:46:33.28 & 55:49:33.5 & $1074.73\pm0.37$ & $76.40\pm0.74$ & $88.16\pm1.11$ & 13.6 & 1.038 & 1.051 & 0.050 & 1.2 & 43.9 & 7.9 & SDSS J114634.04+554916.6 & 11:46:34.08 & 55:49:17.4 & 1079 & \\
FUMa 113 & 11:45:34.84 & 55:53:12.9 & $1151.46\pm0.08$ & $171.48\pm0.16$ & $186.30\pm0.24$ & 78.6 & 10.061 & 10.124 & 0.084 & 1.6 & 212.6 & 8.9 & NGC 3850 & 11:45:35.55 & 55:53:13.1 & 1159 & \\
FUMa 114 & 11:21:01.15 & 53:10:12.5 & $1156.48\pm0.03$ & $100.63\pm0.06$ & $124.65\pm0.09$ & 635.7 & 57.401 & 57.878 & 0.159 & 3.6 & 711.1 & 9.6 & NGC 3631 & 11:21:02.88 & 53:10:11.0 & 1151 & f\\
FUMa 115 & 11:35:02.73 & 54:50:46.9 & $1182.77\pm0.02$ & $235.78\pm0.04$ & $249.26\pm0.06$ & 472.0 & 79.840 & 79.992 & 0.157 & 2.7 & 861.4 & 9.8 & NGC 3733 & 11:35:01.65 & 54:51:02.1 & 1184 & f\\
FUMa 116 & 12:01:01.26 & 55:01:25.8 & $1110.43\pm0.34$ & $95.63\pm0.68$ & $107.53\pm1.02$ & 14.0 & 1.109 & 1.148 & 0.046 & 1.1 & 48.1 & 7.9 & MCG +09-20-063 & 12:01:00.36 & 55:01:32.9 & 1133 & n\\
FUMa 117 & 12:09:30.11 & 53:06:11.2 & $1165.22\pm0.08$ & $178.28\pm0.16$ & $194.09\pm0.24$ & 107.2 & 15.954 & 16.038 & 0.115 & 2.2 & 244.3 & 9.1 & NGC 4142 & 12:09:30.16 & 53:06:17.0 & 1158 & f\\
FUMa 118 & 12:09:56.00 & 42:29:51.7 & $1094.43\pm1.43$ & $126.78\pm2.86$ & $170.20\pm4.29$ & 3.2 & 0.411 & 0.315 & 0.037 & 0.8 & 21.6 & 7.5 & NGC 4143 & 12:09:36.07 & 42:32:03.2 & 946 & f\\
FUMa 119 & 11:59:58.55 & 49:33:55.5 & $1133.72\pm0.25$ & $85.32\pm0.50$ & $100.53\pm0.75$ & 28.1 & 2.395 & 2.394 & 0.068 & 1.6 & 72.3 & 8.2 & MCG +08-22-051 & 11:59:57.69 & 49:33:49.8 & 1124 & \\
FUMa 120 & 12:08:12.98 & 55:44:52.8 & $1121.76\pm1.39$ & $50.47\pm2.78$ & $74.02\pm4.17$ & 11.7 & 0.611 & 0.575 & 0.086 & 2.4 & 16.3 & 7.6 & SDSS J120810.72+554447.2 & 12:08:10.72 & 55:44:47.0 & 1111 & f\\
FUMa 121 & 11:23:56.03 & 52:55:15.8 & $1213.40\pm0.06$ & $197.59\pm0.12$ & $216.88\pm0.18$ & 274.0 & 42.974 & 43.220 & 0.227 & 4.2 & 326.5 & 9.5 & NGC 3657 & 11:23:55.58 & 52:55:15.4 & 1217 & \\
FUMa 122 & 11:50:11.84 & 42:04:14.7 & $1126.61\pm1.26$ & $92.76\pm2.52$ & $126.67\pm3.78$ & 3.0 & 0.242 & 0.253 & 0.018 & 0.5 & 21.6 & 7.2 & UGC 6805 & 11:50:12.29 & 42:04:28.2 & 1158 & \\
FUMa 123 & 12:03:29.77 & 55:03:00.6 & $1118.97\pm1.55$ & $32.37\pm3.10$ & $90.43\pm4.65$ & 4.5 & 0.172 & 0.141 & 0.020 & 0.6 & 23.0 & 7.1 & SDSS J120330.64+550306.0 & 12:03:30.65 & 55:03:06.0 & 1113 & \\
FUMa 124 & 11:51:54.30 & 53:06:09.1 & $1121.16\pm0.54$ & $32.44\pm1.08$ & $50.35\pm1.62$ & 9.0 & 0.313 & 0.304 & 0.024 & 0.7 & 36.6 & 7.3 & SDSS J115153.66+530558.2 & 11:51:53.66 & 53:05:58.2 & 1124 & \\
FUMa 125 & 11:50:56.00 & 48:31:59.2 & $1138.38\pm0.45$ & $68.29\pm0.90$ & $86.65\pm1.35$ & 15.1 & 1.036 & 1.075 & 0.052 & 1.3 & 44.2 & 7.9 & SDSS J115056.11+483153.6 & 11:50:56.17 & 48:31:53.6 & 1127 & \\
FUMa 126 & 11:38:52.48 & 43:09:53.5 & $1197.49\pm0.15$ & $160.18\pm0.30$ & $179.13\pm0.45$ & 43.8 & 6.083 & 6.135 & 0.082 & 1.6 & 133.7 & 8.6 & UGC 6611 & 11:38:51.46 & 43:09:52.6 & 1192 & \\
FUMa 127 & 11:58:51.69 & 45:43:50.8 & $1149.71\pm0.10$ & $40.54\pm0.20$ & $58.91\pm0.30$ & 151.5 & 6.345 & 6.499 & 0.080 & 2.2 & 205.8 & 8.7 & UGCA 259 & 11:58:52.66 & 45:44:03.1 & 1154 & \\
FUMa 128 & 11:32:03.73 & 53:44:09.7 & $1136.21\pm1.47$ & $22.11\pm2.94$ & $51.70\pm4.41$ & 7.0 & 0.186 & 0.168 & 0.035 & 1.0 & 17.3 & 7.1 & SDSS J113202.25+534418.2 & 11:32:02.31 & 53:44:18.0 & 1138  & n\\
FUMa 129 & 11:59:33.84 & 54:13:34.2 & $1141.24\pm0.57$ & $19.83\pm1.14$ & $31.12\pm1.71$ & 17.7 & 0.384 & 0.338 & 0.048 & 1.4 & 27.8 & 7.4 & SDSS J115934.75+541317.9 & 11:59:34.75 & 54:13:17.9 & 1155 & n\\
FUMa 130 & 12:12:56.44 & 53:27:45.3 & $1156.05\pm0.72$ & $34.02\pm1.44$ & $50.11\pm2.16$ & 9.7 & 0.344 & 0.353 & 0.036 & 1.0 & 26.0 & 7.4 & SBS 1210+537A & 12:12:55.89 & 53:27:38.2 & 1155 & f\\
FUMa 131 & 12:17:58.53 & 46:55:56.2 & $1153.60\pm1.37$ & $50.29\pm2.74$ & $68.13\pm4.11$ & 4.1 & 0.197 & 0.174 & 0.031 & 0.9 & 14.4 & 7.1 & SDSS J121759.40+465634.3 & 12:17:59.40 & 46:56:34.3 & 1149 & f\\
FUMa 132 & 11:56:16.00 & 51:16:58.9 & $1160.05\pm0.86$ & $29.45\pm1.72$ & $36.32\pm2.58$ & 10.7 & 0.277 & 0.335 & 0.057 & 1.6 & 14.4 & 7.3 & SDSS J115616.24+511706.9 & 11:56:16.30 & 51:17:07.0 & 1159 & n\\
FUMa 133 & 11:59:59.56 & 44:42:57.9 & $1171.20\pm0.41$ & $52.51\pm0.82$ & $66.93\pm1.23$ & 16.8 & 0.856 & 0.901 & 0.046 & 1.2 & 43.5 & 7.8 & PGC 166118 & 11:59:58.73 & 44:43:05.5 & 1168 & \\
FUMa 134 & 11:39:32.76 & 43:24:41.8 & $1159.09\pm1.05$ & $38.04\pm2.10$ & $44.79\pm3.15$ & 3.9 & 0.147 & 0.187 & 0.032 & 0.9 & 11.6 & 7.0 & SDSS J113930.28+432428.5 & 11:39:30.32 & 43:24:27.8 & 1153 & n\\
FUMa 135 & 11:51:27.58 & 49:47:21.9 & $1197.07\pm0.59$ & $41.88\pm1.18$ & $55.15\pm1.77$ & 17.4 & 0.704 & 0.742 & 0.062 & 1.7 & 28.9 & 7.7 & SDSS J115126.75+494734.4 & 11:51:26.76 & 49:47:34.4 & 1196 & n\\
\hline
\multicolumn{18}{l}{Confused pair} \\
\hline
FUMa 136 & 11:58:47.72 & 42:43:29.8 & $727.68\pm0.09$ & $352.07\pm0.18$ & $384.65\pm0.27$ & 128.8 & 34.085 & 34.301 & 0.178 & 2.7 & 307.0 & 9.4 & UGC 6973 - UGC 6962 &   &   &   & \\
FUMa 137 & 11:37:44.93 & 47:53:43.2 & $740.25\pm0.07$ & $228.96\pm0.14$ & $257.30\pm0.21$ & 406.2 & 69.447 & 69.690 & 0.334 & 5.9 & 351.5 & 9.7 & NGC 3769 - NGC 3769A &   &   &   & f\\
FUMa 138 & 11:48:39.77 & 48:41:40.5 & $968.95\pm0.05$ & $262.84\pm0.10$ & $307.79\pm0.15$ & 432.5 & 113.673 & 113.535 & 0.320 & 5.4 & 588.3 & 9.9 & NGC 3893 - NGC 3896 &   &   &   & f\\
\hline
\multicolumn{18}{l}{Confused small group} \\
\hline
FUMa 139 & 12:06:47.40 & 43:00:37.5 & $809.88\pm0.20$ & $217.24\pm0.40$ & $260.05\pm0.60$ & 270.3 & 51.018 & 50.927 & 0.580 & 10.4 & 150.5 & 9.6 & NGC 4111 group &  &  &  & f\\
FUMa 140 & 11:58:52.29 & 50:51:32.8 & $935.06\pm0.14$ & $64.60\pm0.28$ & $110.56\pm0.42$ & 329.9 & 24.941 & 27.723 & 0.305 & 6.0 & 233.3 & 9.3 & NGC 4026 group &  &  &  & \\
FUMa 141 & 11:57:36.81 & 53:20:59.8 & $1045.87\pm0.05$ & $469.60\pm0.10$ & $478.91\pm0.15$ & 253.2 & 118.898 & 119.682 & 0.635 & 8.5 & 293.2 & 9.9 & NGC 3992 group &  &  &  & f\\
\hline
\multicolumn{18}{l}{Galaxy identified by optical comparison} \\
\hline
FUMa 142 & 11:51:11.85 & 52:01:16.9 & $967.42\pm1.55$ & $28.03\pm3.10$ & $31.81\pm4.65$ & 2.3 & 0.050 & 0.060 & 0.023 & 0.7 & 5.9 & 6.6 & NGC 3931 & 11:51:13.46 & 52:00:03.1 & 914 & \\
FUMa 143 & 11:47:52.91 & 53:50:19.6 & $1090.46\pm1.40$ & $17.12\pm2.80$ & $36.22\pm4.20$ & 5.6 & 0.137 & 0.103 & 0.036 & 1.0 & 14.6 & 7.0 & SDSS J114751.35+535047.9 & 11:47:51.43 & 53:50:48.3 & 1020 & \\
FUMa 144 & 12:00:11.89 & 53:30:07.5 & $1054.62\pm0.91$ & $3.36\pm1.82$ & $33.91\pm2.73$ & 11.3 & 0.086 & 0.059 & 0.021 & 0.7 & 28.5 & 6.8 & 2MASX J11595620+5329451 & 11:59:56.27 & 53:29:44.5 & 1066 & \\
FUMa 145 & 12:15:53.05 & 47:30:38.2 & $687.92\pm1.40$ & $19.27\pm2.80$ & $34.46\pm4.20$ & 2.3 & 0.052 & 0.058 & 0.011 & 0.4 & 13.1 & 6.6 & SDSS J121551.55+473016.8 & 12:15:51.54 & 47:30:17.1 & 655 & n\\
FUMa 146 & 12:31:14.97 & 44:45:26.5 & $962.73\pm1.63$ & $29.73\pm3.26$ & $32.50\pm4.89$ & 0.8 & 0.015 & 0.010 & 0.006 & 0.3 & 4.8 & 6.0 & SDSS J123106.05+444449.1 & 12:31:06.07 & 44:44:49.2 & 948 & n\\
\enddata
\end{deluxetable*}
\end{longrotatetable}}

\setlength{\tabcolsep}{1.3mm}{
\begin{longrotatetable}
\begin{deluxetable*}{lccccccccccccccccc}
\label{tab:UMa catalog2}
\tabletypesize{\tiny}
\centerwidetable
\tablecaption{FUMaS H\,{\footnotesize I} source without known-redshift optical counterpart catalog.}
\tablehead{\colhead{[1]} & \colhead{[2]} & \colhead{[3]} & \colhead{[4]} & \colhead{[5]} & \colhead{[6]} & \colhead{[7]} & \colhead{[8]} & \colhead{[9]} & \colhead{[10]} & \colhead{[11]} & \colhead{[12]} & \colhead{[13]} & \colhead{[14]} & \colhead{[15]} & \colhead{[16]} & \colhead{[17]} & \colhead{[18]} \\ 
\colhead{ID} & \colhead{Ra} & \colhead{Dec} & \colhead{$V_\mathrm{helio}\pm\sigma_\mathrm{v}$} & \colhead{$W_{50}\pm\sigma_{50}$} & \colhead{$W_{20}\pm\sigma_{20}$} & \colhead{$F_\mathrm{peak}$} & \colhead{$S_\mathrm{bf}$} & \colhead{$S_\mathrm{sum}$} & \colhead{$\sigma_\mathrm{s}$} & \colhead{$\sigma_\mathrm{rms}$} & \colhead{SNR} & \colhead{log$M$} & \colhead{Opt name} & \colhead{Opt ra} & \colhead{Opt dec} & \colhead{Opt v} & \colhead{Flags} \\ 
\colhead{} & \colhead{(hms)} & \colhead{(dms)} & \colhead{(km~s$^{-1}$)} & \colhead{(km~s$^{-1}$)} & \colhead{(km~s$^{-1}$)} & \colhead{(mJy)} & \colhead{(Jy~km~s$^{-1}$)} & \colhead{(Jy~km~s$^{-1}$)} & \colhead{(Jy~km~s$^{-1}$)} & \colhead{(mJy)} & \colhead{} & \colhead{(M$_{\sun}$)} & \colhead{} & \colhead{(hms)} & \colhead{(dms)} & \colhead{(km~s$^{-1}$)} & \colhead{} \\ }
\startdata
FUMa 147 & 11:58:47.30 & 55:35:37.7 & $1090.18\pm1.55$ & $31.22\pm3.10$ & $50.93\pm4.65$ & 6.3 & 0.206 & 0.197 & 0.030 & 1.2 & 13.4 & 7.2 & MATLAS 1137 & 11:58:49.56 & 55:35:15.1 &   & n\\
FUMa 148 & 11:49:52.76 & 46:31:45.6 & $713.99\pm1.04$ & $30.30\pm2.08$ & $39.09\pm3.12$ & 6.0 & 0.148 & 0.134 & 0.030 & 0.9 & 13.4 & 7.0 & SDSS J114953.15+463058.9 & 11:49:53.15 & 46:30:58.9 &   & n\\
FUMa 149 & 12:09:17.69 & 43:56:38.9 & $893.44\pm0.60$ & $16.52\pm1.20$ & $30.24\pm1.80$ & 8.1 & 0.148 & 0.194 & 0.021 & 0.6 & 29.1 & 7.0 &  &  &  &   & cf\\
FUMa 150 & 11:32:10.32 & 53:27:18.2 & $717.92\pm1.12$ & $28.09\pm2.24$ & $36.29\pm3.36$ & 6.4 & 0.179 & 0.196 & 0.043 & 1.3 & 12.0 & 7.1 &  &  &  &   & nc\\
FUMa 151 & 11:40:34.19 & 48:15:23.5 & $732.14\pm0.73$ & $20.48\pm1.46$ & $37.85\pm2.19$ & 12.0 & 0.331 & 0.283 & 0.042 & 1.2 & 26.6 & 7.4 & SMDG J1140318+481531 & 11:40:31.53 & 48:15:32.9 &   & n\\
FUMa 152 & 11:53:12.89 & 48:11:01.6 & $759.74\pm0.63$ & $19.20\pm1.26$ & $34.78\pm1.89$ & 17.7 & 0.393 & 0.433 & 0.046 & 1.4 & 29.5 & 7.4 & SDSS J115311.14+481119.2 & 11:53:11.14 & 48:11:19.3 &   & n\\
FUMa 153 & 11:39:03.60 & 45:57:15.0 & $785.78\pm0.59$ & $19.60\pm1.18$ & $28.82\pm1.77$ & 6.9 & 0.141 & 0.150 & 0.020 & 0.6 & 24.1 & 7.0 & SDSS J113901.68+455721.6 & 11:39:01.68 & 45:57:21.6 &   & n\\
FUMa 154 & 11:52:00.36 & 52:33:02.9 & $1044.34\pm1.34$ & $25.40\pm2.68$ & $36.74\pm4.02$ & 7.1 & 0.168 & 0.139 & 0.053 & 1.3 & 11.8 & 7.1 &  &  &  &   & nc\\
FUMa 155 & 11:44:44.02 & 45:00:59.5 & $817.48\pm1.23$ & $17.34\pm2.46$ & $30.90\pm3.69$ & 5.6 & 0.128 & 0.123 & 0.032 & 1.0 & 14.1 & 7.0 & SMDG J1144447+450018 & 11:44:44.70 & 45:00:17.4 &   & n\\
FUMa 156 & 11:29:56.85 & 52:52:39.6 & $861.62\pm0.72$ & $26.77\pm1.44$ & $36.13\pm2.16$ & 7.3 & 0.198 & 0.257 & 0.029 & 0.9 & 20.0 & 7.1 & SMDG J1129572+525248 & 11:29:57.05 & 52:52:47.7 &   & n\\
FUMa 157 & 12:30:06.11 & 50:05:28.4 & $885.37\pm0.82$ & $43.24\pm1.64$ & $67.80\pm2.46$ & 7.1 & 0.330 & 0.385 & 0.029 & 0.8 & 28.2 & 7.4 & SDSS J123006.79+500525.8 & 12:30:06.80 & 50:05:25.8 &   & n\\
FUMa 158 & 11:58:23.91 & 55:04:15.4 & $878.62\pm1.30$ & $32.85\pm2.60$ & $50.68\pm3.90$ & 3.9 & 0.138 & 0.159 & 0.025 & 0.7 & 15.2 & 7.0 & SDSS J115824.37+550418.2 & 11:58:24.37 & 55:04:18.2 &   & n\\
FUMa 159 & 12:21:38.49 & 52:23:03.1 & $881.80\pm1.03$ & $26.69\pm2.06$ & $36.31\pm3.09$ & 5.2 & 0.140 & 0.131 & 0.030 & 0.9 & 14.1 & 7.0 & SDSS J122139.96+522317.9 & 12:21:39.97 & 52:23:18.0 &   & n\\
FUMa 160 & 12:10:12.57 & 43:22:59.6 & $880.66\pm1.22$ & $27.48\pm2.44$ & $47.00\pm3.66$ & 4.5 & 0.145 & 0.141 & 0.025 & 0.7 & 16.9 & 7.0 &  &  &  &   & nc\\
FUMa 161 & 11:54:02.06 & 47:31:14.1 & $937.59\pm0.77$ & $27.08\pm1.54$ & $55.67\pm2.31$ & 12.1 & 0.394 & 0.402 & 0.036 & 1.1 & 32.4 & 7.4 & SDSS J115400.21+473056.0 & 11:54:00.21 & 47:30:56.0 &   & n\\
FUMa 162 & 12:09:27.41 & 43:11:56.2 & $945.19\pm0.46$ & $61.82\pm0.92$ & $71.84\pm1.38$ & 12.3 & 0.760 & 0.755 & 0.044 & 1.4 & 32.4 & 7.7 &  &  &  &  & cf\\
FUMa 163 & 12:05:02.74 & 43:14:00.0 & $923.44\pm0.70$ & $13.06\pm1.40$ & $21.98\pm2.10$ & 9.1 & 0.132 & 0.126 & 0.029 & 0.9 & 18.7 & 7.0 &  &  &  &   & cf\\
FUMa 164 & 11:53:26.77 & 54:12:02.0 & $955.13\pm0.56$ & $44.24\pm1.12$ & $58.00\pm1.68$ & 12.1 & 0.532 & 0.530 & 0.043 & 1.2 & 31.3 & 7.6 &  &  &  &   & n\\
FUMa 165 & 11:58:28.22 & 48:57:50.3 & $957.80\pm0.78$ & $22.87\pm1.56$ & $35.72\pm2.34$ & 10.0 & 0.247 & 0.278 & 0.037 & 1.1 & 21.6 & 7.2 & SMDG J1158259+485737 & 11:58:25.99 & 48:57:36.3 &   & n\\
FUMa 166 & 12:19:33.66 & 43:47:36.1 & $962.80\pm0.31$ & $20.86\pm0.62$ & $32.95\pm0.93$ & 15.9 & 0.373 & 0.381 & 0.023 & 0.7 & 53.4 & 7.4 & SDSS J121933.46+434706.0 & 12:19:33.46 & 43:47:06.0 &   & f\\
FUMa 167 & 11:55:18.60 & 46:19:05.7 & $977.88\pm0.77$ & $27.52\pm1.54$ & $42.34\pm2.31$ & 9.1 & 0.268 & 0.274 & 0.033 & 1.0 & 23.3 & 7.3 & SMDG J1155176+461858 & 11:55:17.56 & 46:18:58.2 &   & n\\ 
FUMa 168 & 12:08:02.09 & 50:53:57.0 & $1026.26\pm0.85$ & $38.18\pm1.70$ & $53.16\pm2.55$ & 7.5 & 0.293 & 0.276 & 0.036 & 1.0 & 21.4 & 7.3 & SDSS J120804.62+505355.1 & 12:08:04.62 & 50:53:55.1 &   & f\\
FUMa 169 & 11:40:02.96 & 49:05:29.6 & $1018.72\pm0.83$ & $27.80\pm1.66$ & $40.62\pm2.49$ & 7.1 & 0.205 & 0.213 & 0.031 & 0.9 & 20.1 & 7.2 & SDSS J114000.43+490606.6 & 11:40:00.43 & 49:06:06.6 &   & n\\
FUMa 170 & 12:05:31.42 & 51:33:08.8 & $1020.61\pm0.64$ & $22.51\pm1.28$ & $24.82\pm1.92$ & 7.2 & 0.151 & 0.163 & 0.043 & 1.3 & 11.1 & 7.0 & SDSS J120536.49+513259.7 & 12:05:36.49 & 51:32:59.7 &   & f\\
FUMa 171 & 11:46:56.16 & 48:04:47.3 & $1042.86\pm0.77$ & $68.10\pm1.54$ & $75.82\pm2.31$ & 5.1 & 0.257 & 0.261 & 0.030 & 0.8 & 16.8 & 7.3 & SDSS J114649.67+480530.5 & 11:46:49.67 & 48:05:30.5 &   & n\\
FUMa 172 & 12:27:33.72 & 51:02:12.4 & $1049.81\pm0.64$ & $30.84\pm1.28$ & $48.57\pm1.92$ & 13.3 & 0.452 & 0.472 & 0.042 & 1.2 & 31.0 & 7.5 & SDSS J122731.01+510131.0 & 12:27:31.01 & 51:01:31.0 &   & f\\
FUMa 173 & 11:50:33.99 & 49:06:52.0 & $1042.54\pm1.58$ & $30.36\pm3.16$ & $35.53\pm4.74$ & 3.5 & 0.088 & 0.068 & 0.034 & 1.1 & 6.7 & 6.8 & SDSS J115034.34+490650.1 & 11:50:34.34 & 49:06:50.2 &   & n\\
FUMa 174 & 11:59:58.43 & 55:53:56.4 & $1062.80\pm1.15$ & $17.19\pm2.30$ & $68.12\pm3.45$ & 14.7 & 0.319 & 0.286 & 0.040 & 1.2 & 29.1 & 7.4 &  &  &  &   & n\\
FUMa 175 & 11:59:30.18 & 55:49:54.5 & $1089.53\pm0.98$ & $16.70\pm1.96$ & $29.31\pm2.94$ & 6.3 & 0.123 & 0.114 & 0.027 & 0.8 & 16.9 & 6.9 & SDSS J115929.42+554946.3 & 11:59:29.42 & 55:49:46.3 &   & n\\
FUMa 176 & 11:52:23.07 & 52:20:53.2 & $1117.84\pm0.68$ & $17.05\pm1.36$ & $42.95\pm2.04$ & 11.3 & 0.249 & 0.254 & 0.026 & 0.8 & 35.1 & 7.2 &  &  &  &   & nc\\
FUMa 177 & 11:53:02.09 & 48:48:52.3 & $1152.73\pm0.75$ & $31.14\pm1.50$ & $39.40\pm2.25$ & 9.3 & 0.239 & 0.248 & 0.037 & 1.1 & 17.9 & 7.2 & SMDG J1153036+484844 & 11:53:03.54 & 48:48:44.9 &   & n\\
FUMa 178 & 11:54:29.19 & 53:20:58.3 & $1165.02\pm1.45$ & $15.91\pm2.90$ & $42.07\pm4.35$ & 5.9 & 0.123 & 0.102 & 0.028 & 0.8 & 16.6 & 6.9 & SMDG J1154306+532048 & 11:54:30.40 & 53:20:49.0 &   & n\\
\enddata
\end{deluxetable*}
\end{longrotatetable}}

\subsection{Comparison with Previous Catalogs}
\label{sec:compare catalog}

The first large UMa catalog was built by \citet{Tully1996}, containing 79 galaxies, and 70 have H\,{\footnotesize I} detections.
Compared to these 79 galaxies, we detected H\,{\footnotesize I} signals for 76 of these sources, except for 1136+46, NGC 3990, and NGC 4346. 
Note that UGC 7129 has a projected distance of 7.53 degrees from the center of the UMa supergroup, which is not included in our catalog. 
Due to the large gas disks and low resolution of radio telescopes compared to optical instruments, H\,{\footnotesize I} sources are subject to confusion, i.e., one H\,{\footnotesize I} source corresponds to more than one optical source.
In particular, there are six galaxies listed in \citet{Tully1996} in three binary galaxy systems which are marked as confused pairs (UGC 6962/6973, NGC 3769/3769A, and NGC 3893/3896) in our catalog. 
There are four galaxies (NGC 3992, UGC 6923, UGC 6940, and UGC 6969) listed in \citet{Tully1996} in the NGC 3992 small group, two galaxies (NGC 4026 and UGC 6956) in the NGC 4026 small group, and six galaxies (NGC 4111, NGC 4117, NGC 4118, UGC 7089, UGC 7094, and 1203+43) in the NGC 4111 small group. 

To study the statistical properties of the Tully-Fisher relation and the dark-matter halos around galaxies, \citet{Verheijen2001} selected 43 spiral galaxies from the UMa members with suitable inclinations to be observed with the WSRT and studied their H\,{\footnotesize I} properties and dynamics. 
\citet{Busekool2021} conducted a blind survey of the UMa region using the VLA, consisting of 54 cross-patterned points covering 16\% of the area.
The table in the article lists 43 galaxies.
These sources are all in our catalog.
The work of \citet{Wolfinger2013} conducted a blind survey of the UMa region using the 76.2m Lovell telescope at Jodrell Bank Observatory, UK.
This work had a detection range of 480 square degrees covering a portion of the UMa region defined by \citet{Tully1996}. 
From their 166 H\,{\footnotesize I} signals, we have selected 53 sources within the UMa region of this work, including 3 new sources that they have confirmed using the Green Bank Telescope (GBT) follow-up observations. 
These 53 sources all have counterparts in our catalog.

\begin{figure}
	\includegraphics[width=\columnwidth]{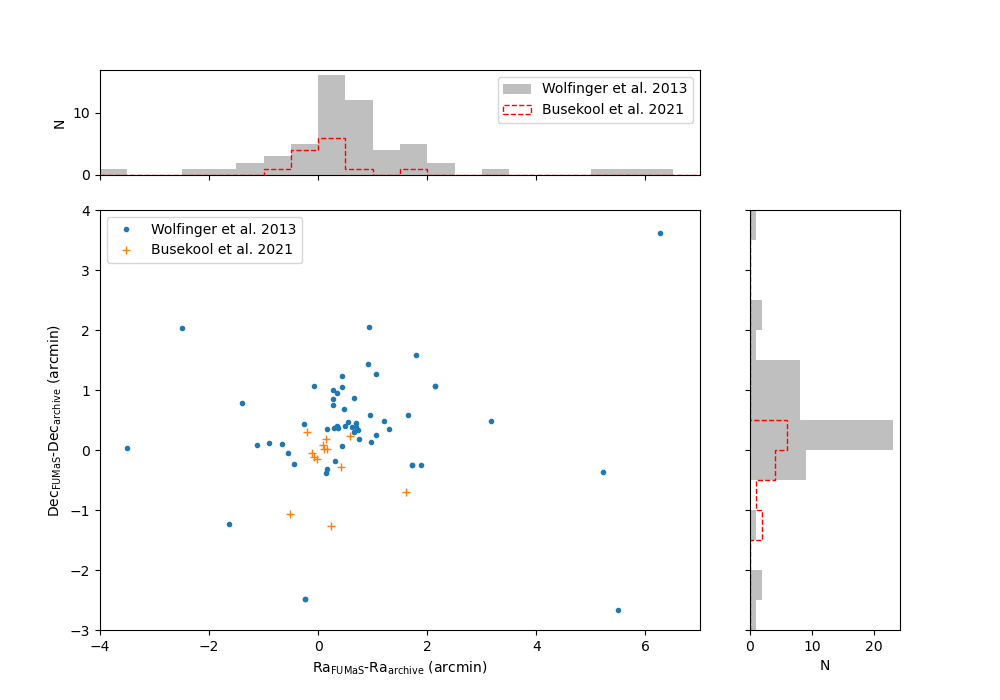}
    \caption{Comparison with H\,{\footnotesize I} sources location information from previous H\,{\footnotesize I} blind survey.
    HIJASS is marked with '.' and \citet{Busekool2021} is marked with '+'. 
    The histogram above shows the distribution of Ra discrepancies, and the distribution of Dec discrepancies is shown on the right.}
    \label{fig:radec_compare}
\end{figure}

\begin{figure}
	\includegraphics[width=\columnwidth]{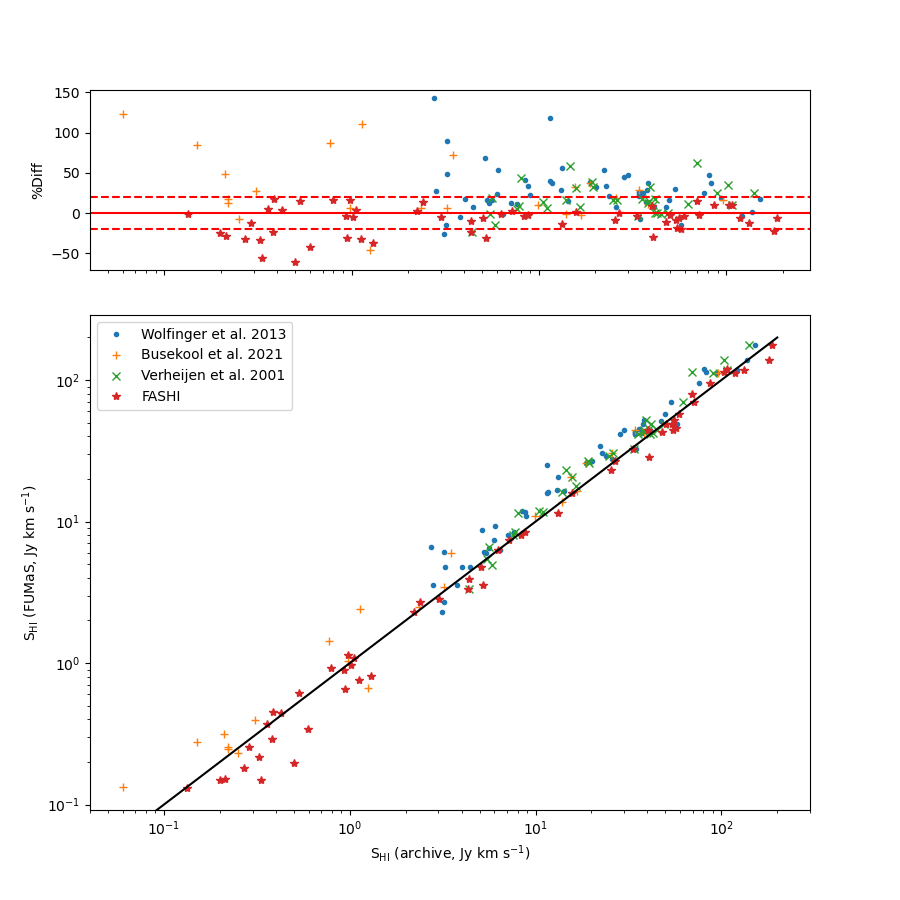}
    \caption{Comparison with flux densities in the literature. 
    The upper plot shows the distribution of percentage difference, with the horizontal straight line being 0 and the two dashed lines representing $\pm$ 20. 
    The lower plot compares flux densities, with the diagonal line representing the case where the two are equal.
    Data in different literature are labeled with different symbols, see figure legend for details.}
    \label{fig:flux_compare}
\end{figure}

\begin{figure}
	\includegraphics[width=\columnwidth]{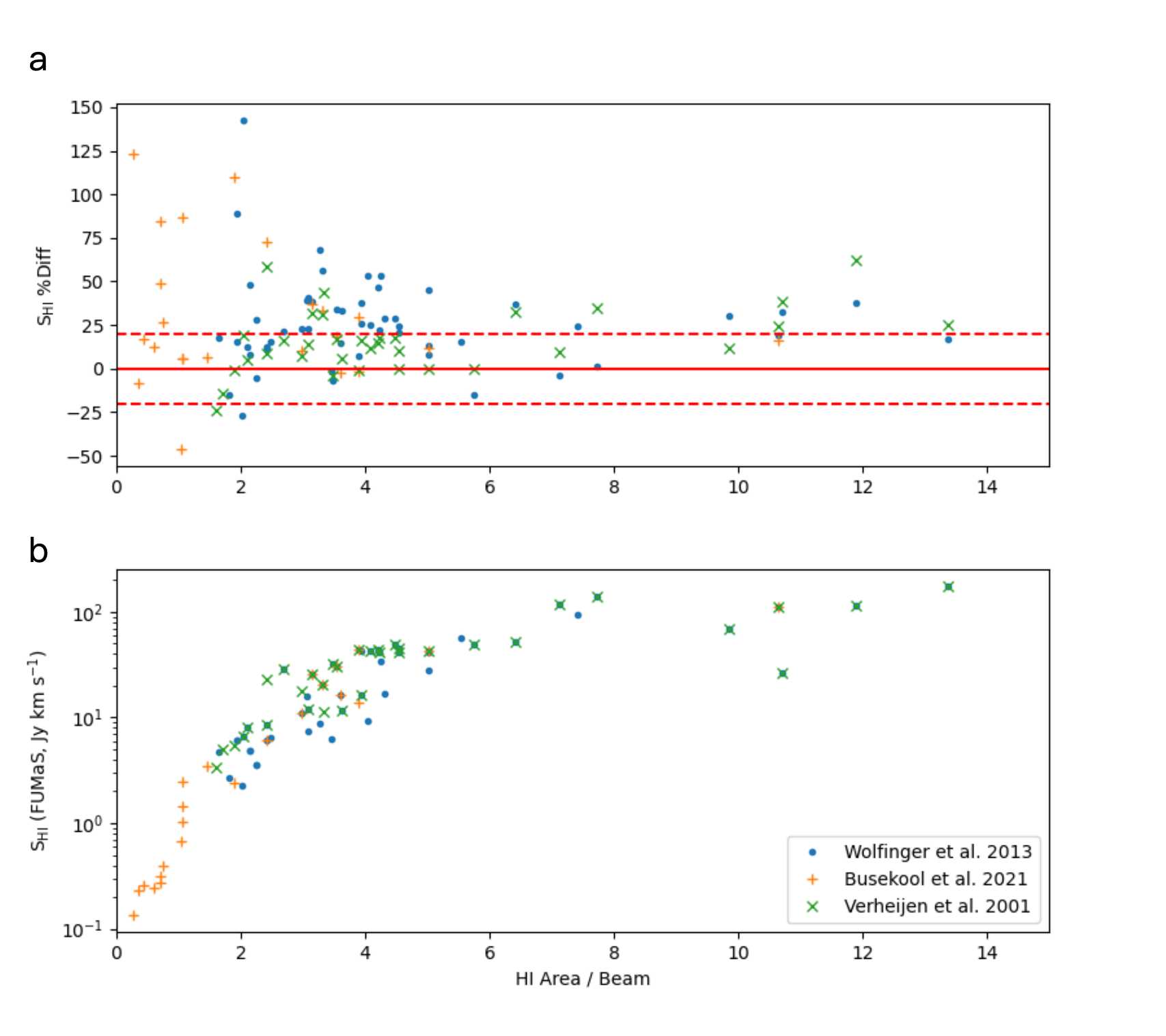}
    \caption{(a) is the relationship between the percentage difference in flux and the Hsize parameter (ratio of the H\,{\footnotesize I} source area to the beam area), and (b) is the relationship between the FUMaS flux and the Hsize parameter.}
    \label{fig:flux_size}
\end{figure}

\begin{figure}
	\includegraphics[width=\columnwidth]{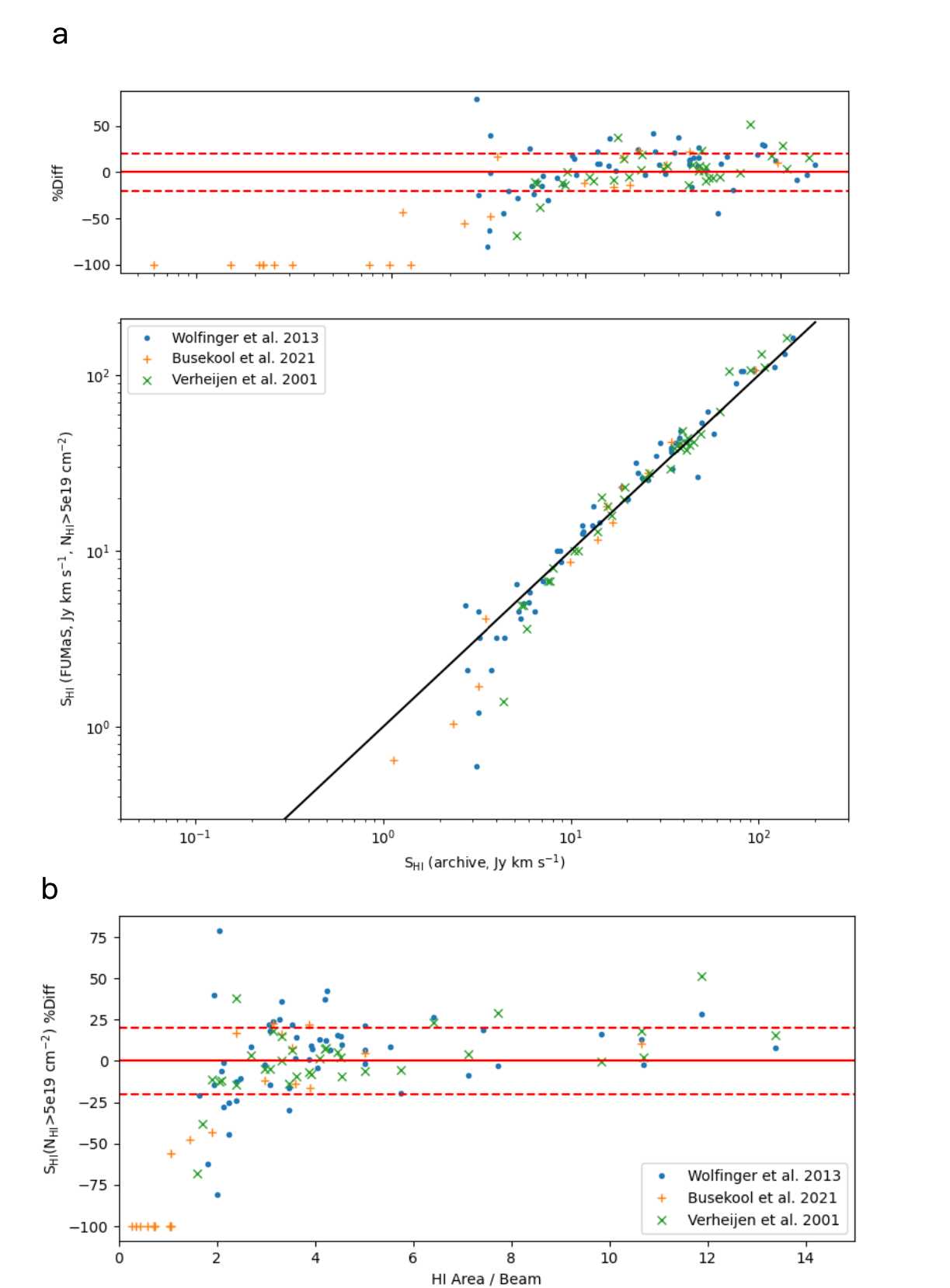}
    \caption{Similar to Figure \ref{fig:flux_compare} and Figure \ref{fig:flux_size}(a), except that the flux densities of the FUMaS detections are for the region of column densities exceeding 5 $\times$ 10$^{19}$ cm$^{-2}$.}
    \label{fig:flux_compare2}
\end{figure}

\begin{figure}
	\includegraphics[width=\columnwidth]{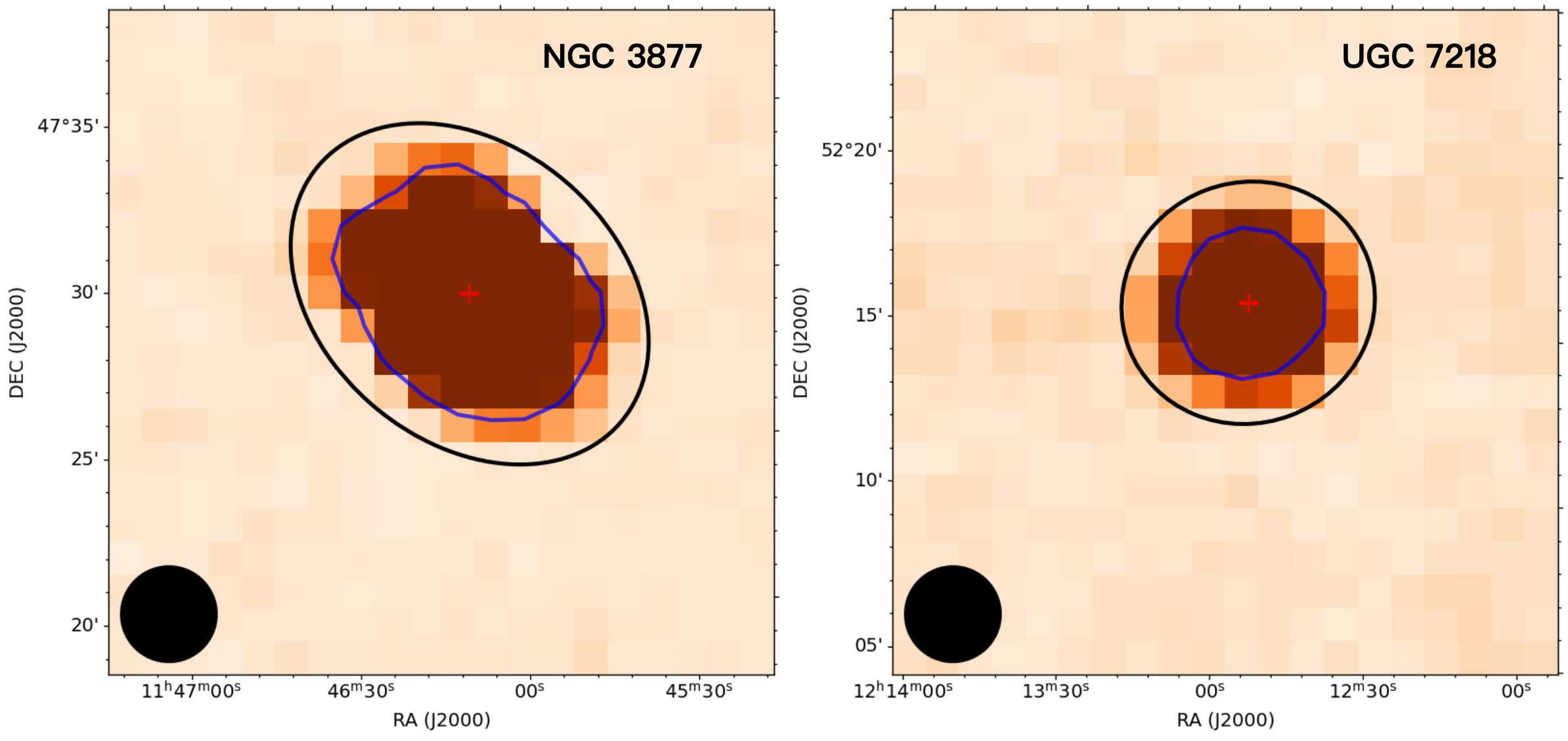}
    \caption{Moment-0 map of NGC 3877 and UGC 7218. 
    The outer black contour is the ellipse size of the source. The inner blue line is the contour with column density 5 $\times$ 10$^{19}$ cm$^{-2}$. 
    The H\,{\footnotesize I} position of the source and the HPBW of FAST are marked in the plot.}
    \label{fig:size}
\end{figure}

\begin{figure}
	\includegraphics[width=\columnwidth]{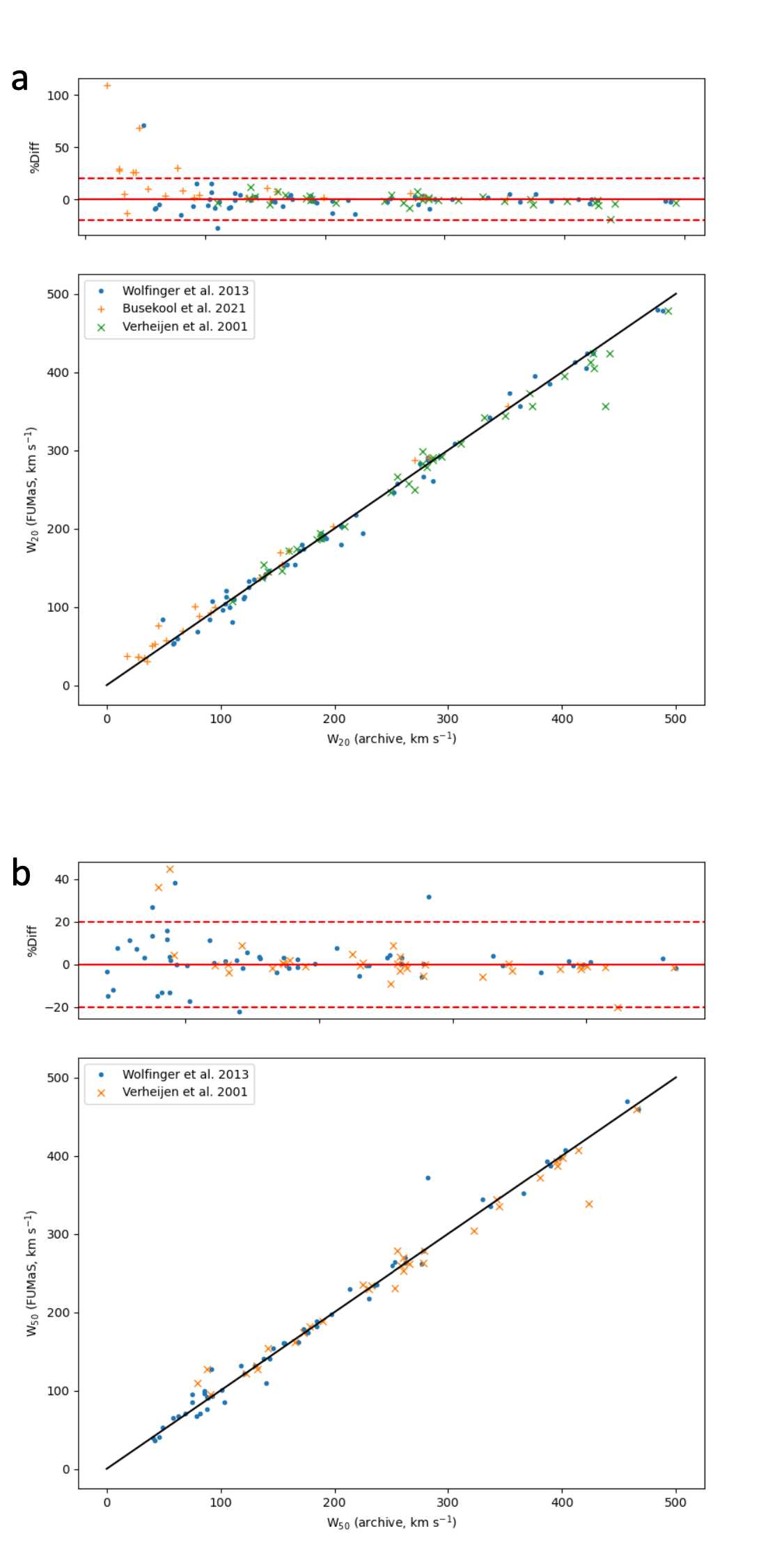}
    \caption{Comparison with the 20\% (a) and 50\% (b) velocity width in the literature.
    The upper plot of each panel shows the percentage distribution of the difference.
    Horizontal lines correspond to 0 and $\pm$ 20, respectively.
    The lower plot is a comparison of the data, and the corresponding symbols for the different literature data are indicated in the legend.
    The diagonal line indicates the consistency of the data.}
    \label{fig:W_compare}
\end{figure}

In Figure \ref{fig:radec_compare} we compare the position of the H\,{\footnotesize I} sources in Table 4 of \citet{Busekool2021}, and the 53 sources in HIJASS, with their counterparts in our catalog.
For the data in \citet{Busekool2021} Table 4, we filter out the sources that are in small groups and outside the UMa region, leaving 13 sources for comparison, denoted by a cross sign in Figure \ref{fig:radec_compare}.
The comparison of the 53 sources of HIJASS is indicated by dots in Figure \ref{fig:radec_compare}.
In the comparison with HIJASS, there are some cases of large positional gaps, which may be due to the low spatial resolution of the single-dish telescope.

To compare flux and velocity widths, we filter out 9 sources in the small groups and select 34 sources in \citet{Verheijen2001}.
Sources in \citet{Busekool2021} with inaccurate information, e.g. those on the edges of observed region, in small groups, and out of the UMa region are filtered, leaving 23 sources for comparison. 
The formula for calculating the percentage difference in Figure \ref{fig:flux_compare} and Figure \ref{fig:W_compare} is as follows:
\begin{equation}
\mathrm{Difference\%} = \frac{\mathrm{DATA}_{\mathrm{FUMaS}}-\mathrm{DATA}_{\mathrm{archive}}}{\mathrm{DATA}_{\mathrm{archive}}}\times100,
\label{eq:quadratic}
\end{equation}
where DATA stands for flux density, 20 percent and 50 percent velocity widths.
Figure \ref{fig:flux_compare} shows the flux densities of FUMaS detections compared to those in the literature.
The results show that the fluxes of FUMaS are systematically high compared to those of either single-dish or interferometer arrays but show good agreement with those of FASHI. 
We note that the FASHI fluxes are also systematically higher by about 10\% than the ALFALFA fluxes for high SNR sources ($>40$). 
\citet{Zhang2024} suggest that this is probably because FAST can recover more diffuse gas than ALFALFA. 
Indeed, \citet{Hoffman2019} pointed out that the ALFALFA pipeline might underestimate the fluxes for bright extended sources by about 10\%. 
The largest differences with the HIJASS fluxes are found in UGC 6773 (142.7\%, inaccuracy of the HIJASS fluxes due to its location near the negative bandpass sidelobes) and the small group NGC 4026 (117.6\%, large amount of diffuse gas in the region of confusion).
To investigate the flux difference with HIJASS, WSRT and VLA observations, we examined the difference as a function of Hsize, as shown in Figure ~\ref{fig:flux_size}(a).
Hsize is the ratio of the H\,{\footnotesize I} source area to the beam area: $\mathrm{Hsize}=(a_{\mathrm{ell}}b_{\mathrm{ell}})/(2.9\arcmin)^2$\citep{Hoffman2019}, where $a_{\mathrm{ell}}$ and $b_{\mathrm{ell}}$ are the half-major and half-minor axes of the elliptical region used to calculate the total flux of the detection, derived from SoFiA. 
For point or small-size sources (Hsize$<3$), the percentage difference fluctuates dramatically because small-size sources have lower fluxes (as can be seen in Figure \ref{fig:flux_size}(b)) close to the noise level. 
Due to the proximity of the UMa region and the lack of previous high-sensitivity surveys, there are not enough comparative data from small-size sources to test the consistency in the flux measurements.
For the extended source (Hsize$>3$), Figure ~\ref{fig:flux_size}(a) clearly shows that the flux of FUMaS is biased by about 20\% compared to other observations.
The excess H\,{\footnotesize I} fluxes measured by FAST could be the results of more diffuse gas with low column density recovered by FAST. 
Such effect has been seen in several cases for FAST observations of extended galaxies \citep[e.g.,][]{Xu2021, Yu2023, Zhou2023, Liu2024}.
The FAST Extended Atlas of Selected Targets Survey (FEASTS) team,  using their own data processing pipeline to derive the H\,{\footnotesize I} fluxes for a sample of 10 extended galaxies,  obtained results similar to ours, which show that the VLA observations miss a median of 23\% of H\,{\footnotesize I} due to interferometry's short-spacing problem and limited sensitivity compared to FAST \citep{Wang2024}.

To further demonstrate this point, we choose to sum up only the FAST fluxes in regions with column densities greater than 5 $\times$ 10$^{19}$ cm$^{-2}$ and make a new comparison plot in Figure \ref{fig:flux_compare2}.
In Figure ~\ref{fig:size}, we show the elliptical size of the H\,{\footnotesize I} source and the regions with column densities greater than 5 $\times$ 10$^{19}$ cm$^{-2}$, using NGC 3877 (extended source, Hsize = 3.1) and UGC 7218 (small-size source, Hsize = 1.6) as examples.
In the results of Figure ~\ref{fig:flux_compare2}(a) we can see that this time we obtain more consistent fluxes, with a median flux percentage difference of only 1.2\% in Figure \ref{fig:flux_compare2}(a), in contrast with the large difference of 17.9\% in Figure \ref{fig:flux_compare}.
In Figure ~\ref{fig:flux_compare2}(b) we can see that the consistency of the flux of the extended source is much more pronounced.
Such agreement proves that the systematic bias in the fluxes is due to the high sensitivity of FAST to recover more diffuse gas with low column densities around the extended sources.
Although this effect is common in single-dish versus interferometer array comparisons, the agreement between HIJASS and \citet{Verheijen2001} fluxes suggests that HIJASS did not detect more diffuse gas, and thus could also miss some low column density gas and resulted in a flux density value lower than that of FUMaS.
Note that the negative flux difference at the small size end is due to their size being close to the spatial resolution of FAST, so the method of removing the fluxes from the low column density regions introduces a significant error. 
In particular, for the newly discovered H\,{\footnotesize I} sources in \citet{Busekool2021}, which are small in size and low in flux, there are no regions with column densities above 5 $\times$ 10$^{19}$ cm$^{-2}$, so the flux percentage difference in Figure ~\ref{fig:flux_compare2} is -100\%. 
In summary, FAST can detect about 20\% of the diffuse H\,{\footnotesize I} around the extended sources compared to other previous observations. 
This is consistent with the findings of FEASTS \citep{Wang2024}, which show that the fractions of diffuse H\,{\footnotesize I} over the total H\,{\footnotesize I} range from 5\% to 55\% with a median value of $\sim $34\% among their sample.
However, it should be caution that our study is limited to the H\,{\footnotesize I} sources in the UMa supergroup, which are in a disturbed environment, thus our conclusion for FAST detecting more diffuse gas is preliminary, and further studies with a large samples are needed to reach a final conclusion.

The velocity width comparisons show good agreement in Figure \ref{fig:W_compare}. 
Most of the data differences are within $\pm$ 20\%.

For the optical counterparts, we compared the sources in \citet{Pak2014}. 
This work confirms 166 galaxies in the UMa supergroup using SDSS-DR7 and NED spectroscopic data. 
They found that while the UMa supergroup is dominated by late-type galaxies, it has a significant number of early-type galaxies.
In the primary comparison, H\,{\footnotesize I} signals were detected for 146 optical galaxies. 
We then reconfirmed the spectra at locations where no signal was detected and identified 4 H\,{\footnotesize I} sources (SDSS J114751.35+535047.9, 2MASX J11595620+5329451, SDSS J121551.55+473016.8 and SDSS J123106.05+444449.1). 
As a result, we detected H\,{\footnotesize I} signals in 150 of the 166 galaxies in the optical catalog, and 16 did not. 

\begin{figure*}
	\includegraphics[width=2\columnwidth]{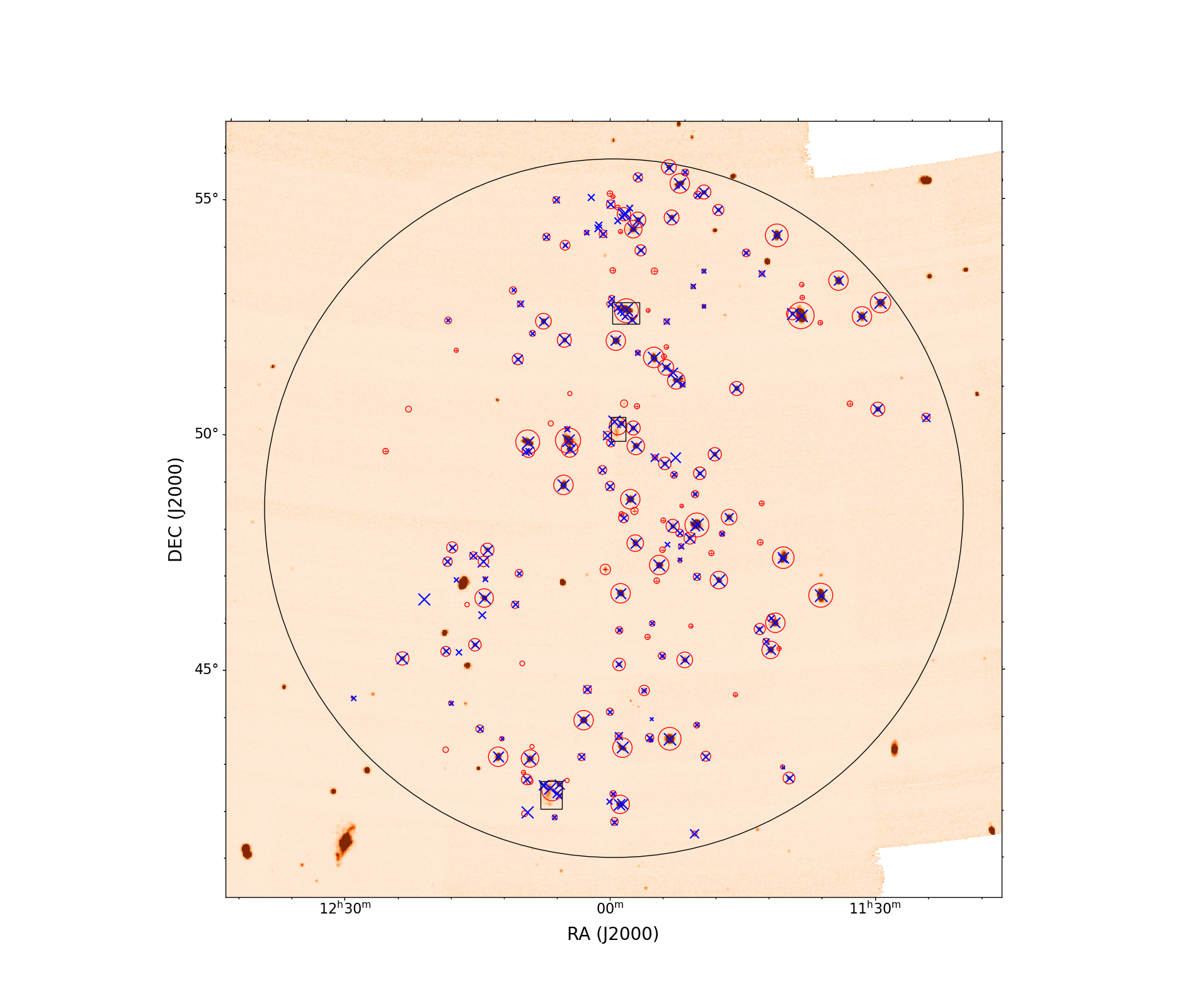}
    \caption{Moment-0 map of the UMa region for the 500-1500 km~s$^{-1}$ integral range.
    The black circle shows the range of the UMa region. 
    Red circles mark the positions of detections in the FUMaS catalog, with the crosses being the first detected H\,{\footnotesize I} sources.
    The blue crosses are optical galaxies in \citet{Pak2014}. 
    Boxes mark the positions of the small groups.
    The size of the marker is proportional to the H\,{\footnotesize I} mass and r magnitude.}
    \label{fig:m0}
\end{figure*}

\begin{figure}
	\includegraphics[width=\columnwidth]{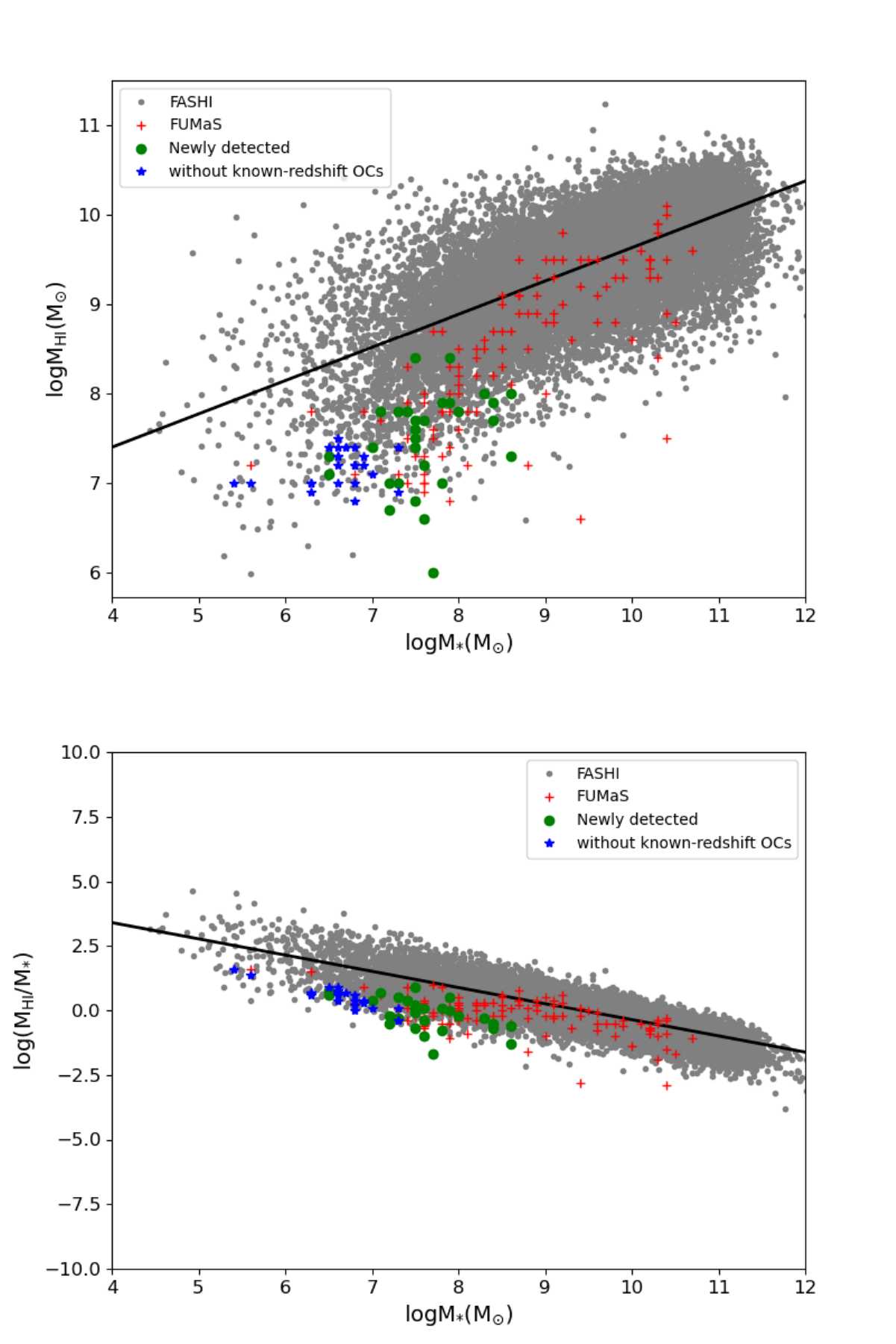}
    \caption{H\,{\footnotesize I} mass and stellar mass distribution of FUMaS detections.
    The top panel is log(M$_{\mathrm{HI}}$) vs log(M$_*$), while the bottom is log(M$_{\mathrm{HI}}$/M$_*$) vs log(M$_*$).
    The grey dots in the background are FASHI data, and the black line indicates its average distribution. 
    The red crosses are previously detected H\,{\footnotesize I} sources; the green dots are new FUMaS detections with known-redshift optical counterparts; and the blue stars are those without known-redshift optical counterparts.
    }
    \label{fig:mass}
\end{figure}

In comparison with previous H\,{\footnotesize I} detection works in the UMa region, new sources detected by blind surveys in both VLA and HIJASS have been included in our catalog, proving the high completeness of our work.
In comparison with the optical work, we have obtained 5 more detections from optically selected positions. 
They are listed at the end of Table \ref{tab:UMa catalog1}, FUMa 142 to FUMa 146.
Finally, our catalog extends to 178 detections, of which 146 have known-redshift optical counterparts and 32 do not.
There are 58 H\,{\footnotesize I} sources published in the first release of FASHI, and they are flagged with 'f' in the catalog. 
In addition to the blind surveys mentioned above, there are a number of optically selected observations that provide information on H\,{\footnotesize I} in the UMa supergroup \citep{Paturel2003, Huchtmeier2007, Courtois2015, vanDriel2016, Poulain2022, O'Neil2023, Chandola2024}. 
For comparison, there are 55 H\,{\footnotesize I} sources in the FUMaS catalog that have been detected for the first time and are flagged with 'n'.
Figure \ref{fig:m0} shows the moment-0 map for the UMa region with an integrated range of 500-1500 km~s$^{-1}$, with detections in our catalog labeled as red hollow circles, optical galaxies in \citet{Pak2014} labeled as blue crosses, and 3 boxes for regions of the small groups.
The size of the marker is proportional to the H\,{\footnotesize I} mass and r magnitude.
The H\,{\footnotesize I} mass and stellar mass distribution of the FUMaS detections are shown in Figure \ref{fig:mass}. 
The stellar mass is calculated by the mass-to-light ratio equation \citep{Bell2003}: $log_{10}(M/L_r)=-0.306+1.097(g-r)$.
As expected, galaxies in the UMa supergroup are influenced by their environment, and the H\,{\footnotesize I} mass of dwarf galaxies is lower relative to the average value of the FASHI global distribution. 
The M$_{\mathrm{HI}}$/M$_*$ ratios for the UMa galaxies are in general lower than that of the main sequence galaxies.
The newly detected H\,{\footnotesize I} sources in FUMaS correspond to low-mass galaxies, and those without known-redshift optical counterparts correspond to even lower-mass galaxies. 
This demonstrates that high sensitivity observations with FAST are crucial in revealing the environmental effects in high density environments. 

\section{Discussion and Results}
\label{sec:discussion and results}

\subsection{H\,{\footnotesize I} sources without known-redshift optical counterparts}
\label{sec:hi noredopt}

\begin{figure*}
	\includegraphics[width=2\columnwidth]{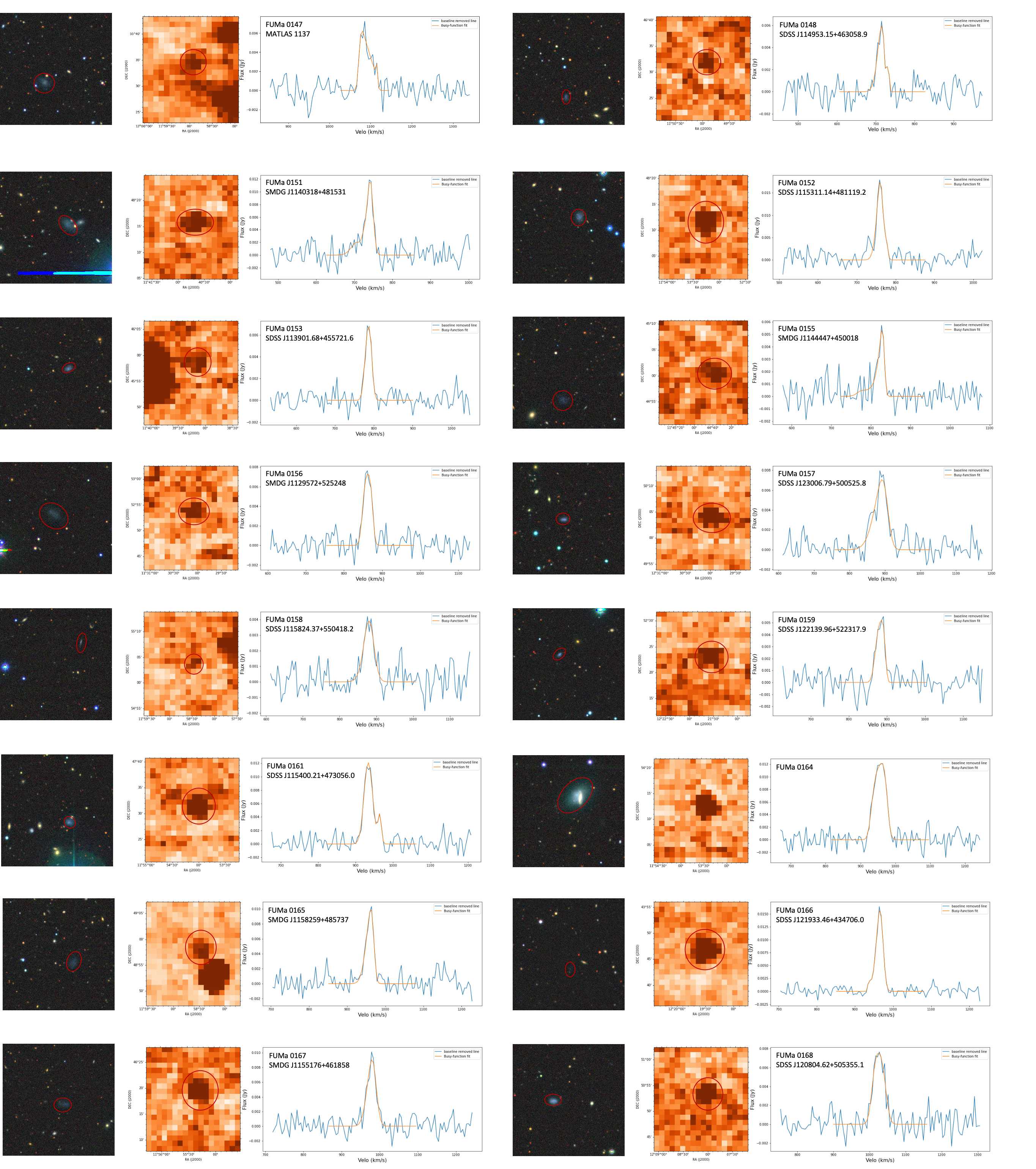}
    \caption{DESI Legacy Surveys DR10 multicolor images, moment-0 maps and profiles for 25 H\,{\footnotesize I} sources with optical counterparts without a redshift. 
    The Legacy survey images are centered on the location of the H\,{\footnotesize I} sources and have a range of about 3$\times$3 arcmin$^2$. 
    The red circles mark the optical counterparts. 
    The range of the H\,{\footnotesize I} sources are indicated in the moment-0 maps. 
    The H\,{\footnotesize I} profile plots contain the names of the detections and optical counterparts, the baseline removed profiles and the busy-function fitted profiles.}
    \label{fig:noredopt1}
\end{figure*}

\begin{figure*}
	\includegraphics[width=2\columnwidth]{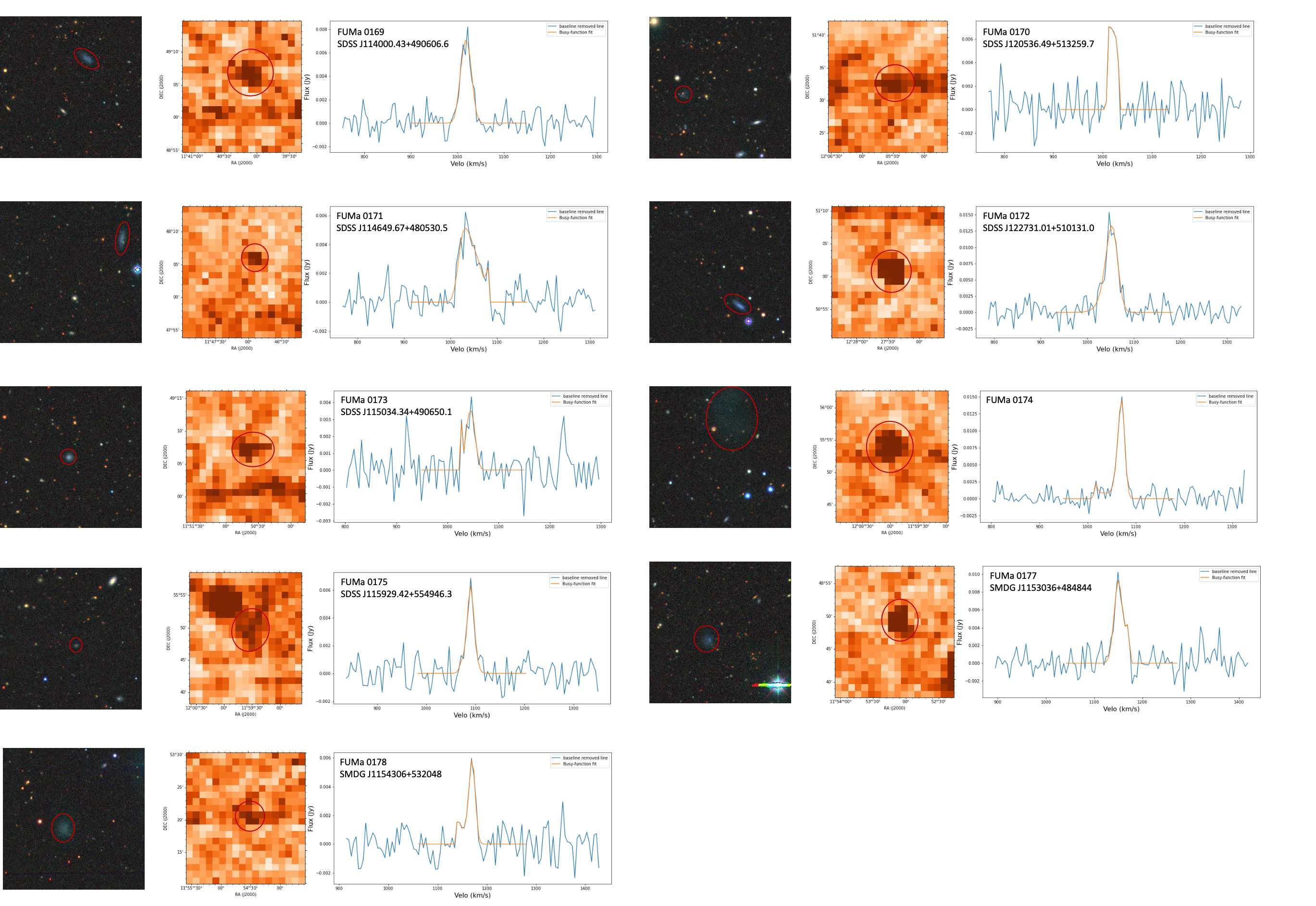}
    \setcounter{figure}{10}
    \caption{\textit{(continued)}}
    \label{fig:noredopt2}
\end{figure*}

Thanks to the high sensitivity of FAST, we can detect more low-mass H\,{\footnotesize I} sources, with the lowest mass of H\,{\footnotesize I} source in the catalog at 10$^{6.0}$ M$_{\sun}$. 
There are 55 H\,{\footnotesize I} sources in our catalog that are new detections, which are labeled as red circles with crosses in Figure \ref{fig:m0}.
They have masses in the range of 10$^{6.0}$-10$^{8.4}$ M$_{\sun}$.
In addition, 32 H\,{\footnotesize I} sources listed in Table \ref{tab:UMa catalog2} have no redshift from optical spectra. 
These sources have masses in the range of 10$^{6.8}$-10$^{7.7}$ M$_{\sun}$, and they may be dwarf galaxies or pure H\,{\footnotesize I} clouds.
For 25 of them, we found possible optical counterparts in optical images.
The optical images are from the Dark Energy Spectroscopic Instrument (DESI) Legacy Surveys\footnote{\url{https://www.legacysurvey.org/viewer}}, and the center of the images is the center of the H\,{\footnotesize I} sources with a size of about 3$\times$3 arcmin$^2$, which corresponds to the HPBW of FAST.
Their multicolor optical images, H\,{\footnotesize I} moment-0 maps, and spectral line profiles are shown in Figure \ref{fig:noredopt1}.  
Seven of these H\,{\footnotesize I} sources (FUMa 151, FUMa 155, FUMa 156, FUMa 165, FUMa 167, FUMa 177 and FUMa 178) correspond to ultra-diffuse galaxies (UDG) in the Systematically Measuring Ultra-diffuse Galaxies (SMUDGes) catalog \citep{Zaritsky2023}.
Sixteen H\,{\footnotesize I} sources appear to have optical counterparts, and we have labeled them in Figure \ref{fig:noredopt1} and listed the related information obtained from NED or SDSS DR16.
FUMa 164 and FUMa 174 also have possible optical counterparts but have not been included in any catalogs.
These two galaxies are first discovered by our H\,{\footnotesize I} survey. 
The UDG galaxy FUMa 164 appears to be confused with a bright galaxies MCG +09-20-024 with a redshift of 0.06.
FUMa 174 seems to be a UDG as well, but its surface brightness is very low so it hasn't been noticed before.
The remaining 7 H\,{\footnotesize I} sources do not have obvious optical counterparts in the optical images.
However, these sources are all close to other galaxies with optical counterparts, so they have a high probability of being H\,{\footnotesize I} clouds.
Deep multi-wavelength follow-up observations are needed to reveal the nature of these H\,{\footnotesize I} sources.

\subsection{Small groups}
\label{sec:complex}

Interactions between galaxies are common in high density regions like groups, clusters, and supergroups, and provide important information for studying the formation and evolution of galaxies. 
There are many interacting systems and close pairs in the UMa supergroup. 
Some of the interacting galaxy systems have been mentioned in previous papers, such as NGC 4085/4088, NGC 3718/3729, and the NGC 3998 system \citep{Verheijen2001, Wolfinger2013}.
We then describe in detail the distribution of H\,{\footnotesize I} in the three small groups in the catalog, as well as in the NGC 3998 group.

\begin{figure*}
	\includegraphics[width=2\columnwidth]{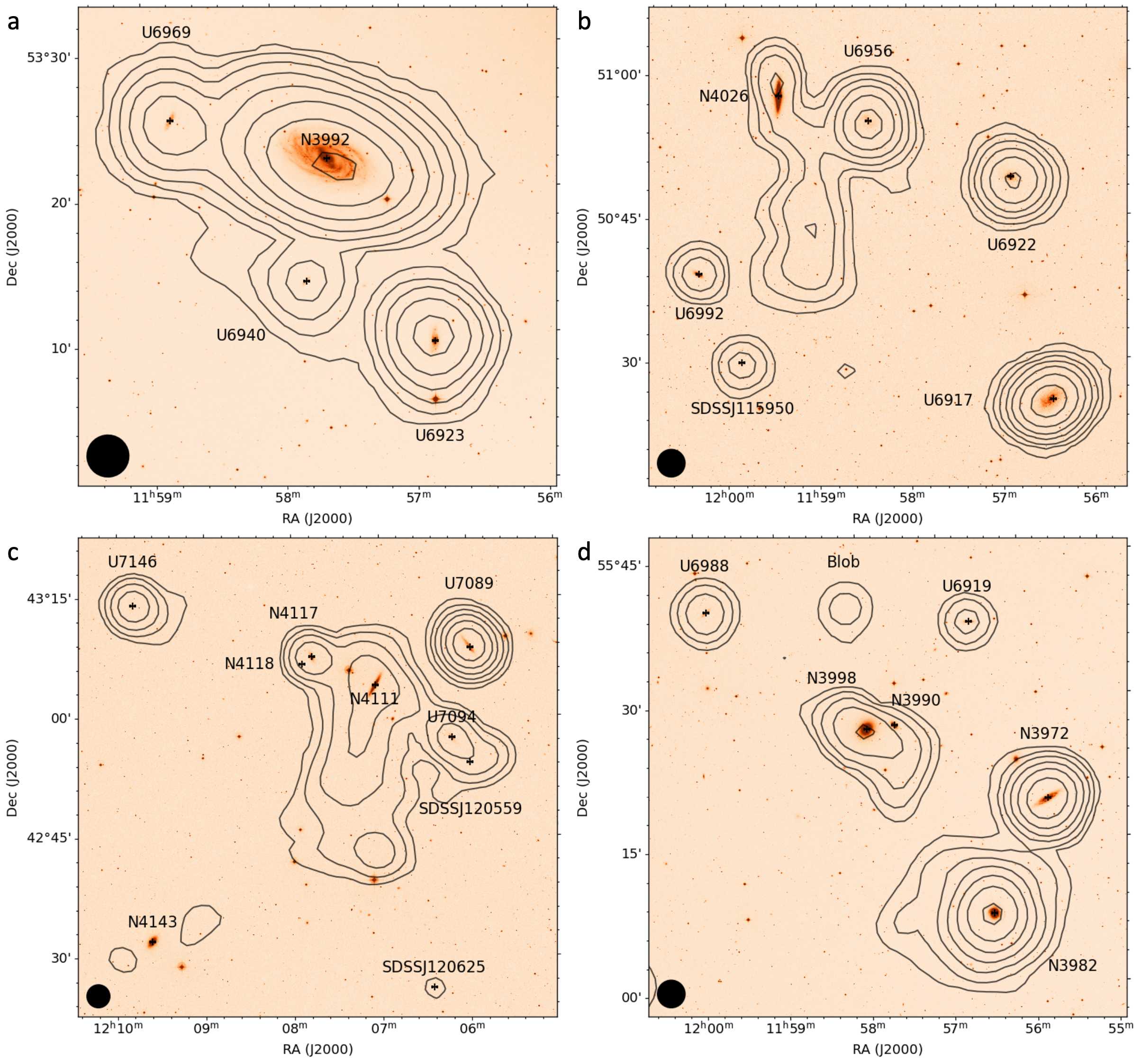}
    \caption{H\,{\footnotesize I} column density contours of 4 small groups overlaid on the DSS B-band optical image.
    Their integrated ranges are 500-1500 km~s$^{-1}$ for NGC 3992 group (a), 600-1400 km~s$^{-1}$ for NGC 4026 group (b), 500-1300 km~s$^{-1}$ for NGC 4111 group (c), and 550-1500 km~s$^{-1}$ for NGC 3998 group (d) respectively
    The contour levels are 6 $\times$ 2$^n$ $\times$ 10$^{18}$ cm$^{-2}$ ($n=0,1,2,...$).
    The optical centers of each galaxy are marked as crosses.
    The HPBW of FAST is indicated at the bottom-left corner of each panel.}
    \label{fig:group}
\end{figure*}

NGC 3992 is a barred spiral galaxy. 
It has three companion galaxies, UGC 6969, UGC 6940 and UGC 6923.
Previous H\,{\footnotesize I} observations by VLA and WSRT have shown that the gas kinematics of NGC 3992 are very regular and that there is a lack of H\,{\footnotesize I} gas in the central bar region. 
The gas disks of the three companion galaxies are as large as the optical disks and do not appear to overlap \citep{Gottesman1984, Verheijen2001, Bottema2002, Bottema2002-2}.
Figure \ref{fig:group}(a) shows the H\,{\footnotesize I} column density contours from FAST observations overlaid on a Digitized Sky Survey (DSS\footnote{\url{https://archive.eso.org/dss/dss}}) b-band image, with an integration range of 500-1500 km~s$^{-1}$ and column density contours starting at 6 $\times$ 10$^{18}$ cm$^{-2}$. 
Due to the high sensitivity of FAST, the four galaxies are surrounded by a large amount of diffuse gas, causing them to lie within a common H\,{\footnotesize I} envelope, and the asymmetry of the gas in the outskirts of UGC 6940 signals the presence of interactions in the group. 
Due to the overlap of the gas disks, we can obtain approximate flux densities and masses for the galaxies, which are given in Table \ref{tab:feature_table}.

NGC 4026 is one of the three brightest lenticular galaxies in the UMa supergroup.
Previous large beam-size single-dish telescopes have obtained a large gaseous envelope containing numerous galaxies in this region, including NGC 4026, UGC 6956, UGC 6922, UGC 6917, and UGC 6992 \citep{Appleton1983, Wolfinger2013}. 
High-resolution interferometer array observations reveal a filament structure south of NGC 4026 \citep{vanDriel1988, Verheijen2001ASPC, Serra2012}.
The column density contours from FAST observations of the NGC 4026 group are shown in Figure \ref{fig:group}(b). 
Due to the higher spatial resolution compared to other single-dish telescopes, our result distinguishes the gas disks of UGC 6917, UGC 6922, UGC 6992, and SDSS J115950.81+502955.3.
The filament extends from the north of NGC 4026's optical center to the south near the SDSS J115950.81+502955.3. 
It has a projected length of about 28\arcmin, corresponding to 142 kpc, and a flux density of about 11.67 Jy, corresponding to a mass of 8.3 $\times$ 10$^{8}$ M$_{\odot}$. 
The approximate flux densities and masses of NGC 4026 and UGC 6956 are given in Table \ref{tab:feature_table}.

NGC 4111 is also one of the three brightest lenticular galaxies. 
It is surrounded by several companion galaxies, with NGC 4117 and NGC 4118 to the east and UGC 7089, UGC 7094, and SDSS J120559.63+425409.1 to the west. 
In the HIJASS observations, they are within a common envelope \citep{Wolfinger2013}.
A little further away are UGC 7146, NGC 4143, and SDSS J120625.35+422604.7. 
WSRT and VLA observations show the filament structure to the south of NGC 4111 \citep{Verheijen2001ASPC, Serra2012, Busekool2021}.
As shown by the FAST observation in Figure \ref{fig:group}(c), the filament extends southwards from NGC 4111 and appears to extend southeastwards to near NGC 4143. 
This filament has a projection of up to 30\arcmin, corresponding to a length of 152 kpc. 
NGC 4143 has no H\,{\footnotesize I} gas in its optical position, but it has two clumps of gas at its southeast and northwest ends, which may represent preexisting gas stripped out by internal or external influences.
The properties of the galaxies in this system are also listed in Table \ref{tab:feature_table}.

The last of the three brightest lenticular galaxies is NGC 3998, which has a disturbed and more fragmented H\,{\footnotesize I} gas distribution \citep{Verheijen2001ASPC, Serra2012}.
The column density contour of the NGC 3998 group is shown in Figure \ref{fig:group}(d), with asymmetries in the gas disks of both NGC 3998 and NGC 3982. 
Although NGC 3990 lies within the gas disk of NGC 3998, there is no obvious dense gas at its optical center, and it is not observed by either the WSRT or the VLA \citep{Verheijen2001ASPC, Serra2012}, so we consider it to be a gas-deficient S0 galaxy.
There is a H\,{\footnotesize I} blob without an optical counterpart just north of NGC 3998, consistent with previous observations \citep{Frank2016}.
Since the galaxies in the NGC 3998 group do not overlap in three dimensions like the three small groups above, they are not included in the catalog as a small group but as individual galaxies.
The northern H\,{\footnotesize I} blob and UGC 6919 are not listed in our catalog due to their high velocities.

\begin{table}
	\centering
	\caption{H\,{\footnotesize I} properties of components in small groups.}
	\label{tab:feature_table}
	\begin{tabular}{lcc} 
		\hline
		Name & Flux (Jy km~s$^{-1}$) & log$M$(M$_{\sun}$)\\
		\hline
        NGC 3992 & 103.911 & 9.9\\
		UGC 6969 & 7.849 & 8.7\\
		UGC 6940 & 2.619 & 8.3\\
        UGC 6923 & 13.144 & 9\\ 
        NGC 4026 & 2.575 & 8.3\\
        UGC 6956 & 14.227 & 9\\
        NGC 4026 filament & 11.67 & 8.9\\
        NGC 4111 & 8.34 & 8.8\\
        NGC 4117 & 2.992 & 8.3\\
        NGC 4118 & 0.732 & 7.7\\
        UGC 7089 & 16.912 & 9.1\\
        UGC 7094 & 5.601 & 8.6\\
        NGC 4111 filament & 16.71 & 9.1\\
		\hline
	\end{tabular}
\end{table}

\section{H\,{\footnotesize I} mass function}
\label{sec:HIMF}

The number density function describing the H\,{\footnotesize I} mass of galaxies can be expressed as:
\begin{equation}
    \phi(M_{\mathrm{HI}})=\frac{dN_\mathrm{gal}}{dVd\mathrm{log}_{10}(M_{\mathrm{HI}})},
	\label{eq:himf}
\end{equation}
where $dN_\mathrm{gal}$ is the number of galaxies with H\,{\footnotesize I} mass in the logarithmic bin $d\mathrm{log}_{10}(M_{\mathrm{HI}})$ within the volume $dV$.
In order to accurately calculate the H\,{\footnotesize I}MF of the UMa supergroup galaxies, it is important to distinguish the mass of individual galaxies and to determine the volume range. 
In the catalog of this paper, the majority of H\,{\footnotesize I} sources correspond to single galaxies, except for three confused pairs and three small groups.
We roughly divided the three small groups into individual galaxies, whose properties are shown in Table \ref{tab:feature_table}. 
For the three confused pairs, it is difficult to distinguish individual galaxy properties in space and velocity, so we divided the H\,{\footnotesize I} masses by their stellar mass ratios.
In addition, seven H\,{\footnotesize I} clouds without optical counterparts and two filaments in the small groups were excluded. 
For the selection of distances and volumes, we used the same strategy as in \citet{Busekool2021}, i.e., the masses of the galaxies were all calculated using an average distance of 17.4 Mpc from the UMa supergroup, and the depth range was based on the Cosmic Flows 2 of 14.7-21.6 Mpc.
Each of these assumptions proved to have little impact on the result of the UMa supergroup H\,{\footnotesize I}MF calculation.
Combined with the projected radius of the UMa region of 7.5 degrees, the volume of the region was calculated to be 123.65 Mpc$^3$ using the formula:
\begin{equation}
    V=\frac{1}{3}\Omega(d_\mathrm{max}^3-d_\mathrm{min}^3).
    \label{eq:v}
\end{equation}

The current calculation of the H\,{\footnotesize I}MF is based on two main methods, the 1/V$_\mathrm{max}$ method \citep{Schmidt1968} and the 2-dimensional stepwise maximum likelihood estimator \citep[2DSWML,][]{Efstathiou1988}. 
The latter can avoid the influence of LSS but will introduce more errors for small sample data. 
For the UMa supergroup, the 1/V$_\mathrm{max}$ method is more appropriate. 
The principle of this method is to calculate the maximum volume V$_\mathrm{max}$ of each galaxy that can be detected. 
In order to calculate the maximum volume, we should first calculate the maximum distance of the galaxies, D$_\mathrm{max}$, at which each galaxy mass can be detected on its noise scale. 
For this noise scale, we chose the same parameter as in SoFiA, 4$\sigma_\mathrm{s}$.
It is worth noting that this distance is limited to 14.7-21.6 Mpc.
If D$_\mathrm{max}$ exceeds 21.6 Mpc, the value 21.6 Mpc will be taken.
Then the maximum volume between 14.7 and D$_\mathrm{max}$ is calculated by equation \ref{eq:v}.
The final fit is performed by the following Schechter function:
\begin{equation}
    \phi(M_{\mathrm{HI}})=ln(10)\phi_*(\frac{M_{\mathrm{HI}}}{M_*})^{\alpha+1}e^{-(\frac{M_{\mathrm{HI}}}{M_*})},
	\label{eq:Schechter}
\end{equation}
where $\phi_*$ is the normalisation constant, $\alpha$ is the low mass slope and $M_*$ is the characteristic mass. 
Due to the low completeness of the low mass detections, we filtered out 8 detections with masses less than 10$^{6.9}$ M$_{\odot}$.
The non-linear least squares (NLLS) method is widely used for fitting.
We binned the samples at 0.2 dex width and fitted using the LMFIT package\footnote{\url{https://lmfit.github.io/lmfit-py/}}.
Still, the selection and width of the bins can affect the results of the fitted parameters, particularly for small sample sizes. 
We therefore carried out another fitting using the modified maximum likelihood (MML) method \citep{Obreschkow2018}, which is a more robust method that does not require the bins to be split and takes the mass error into account.

\begin{figure}
	\includegraphics[width=\columnwidth]{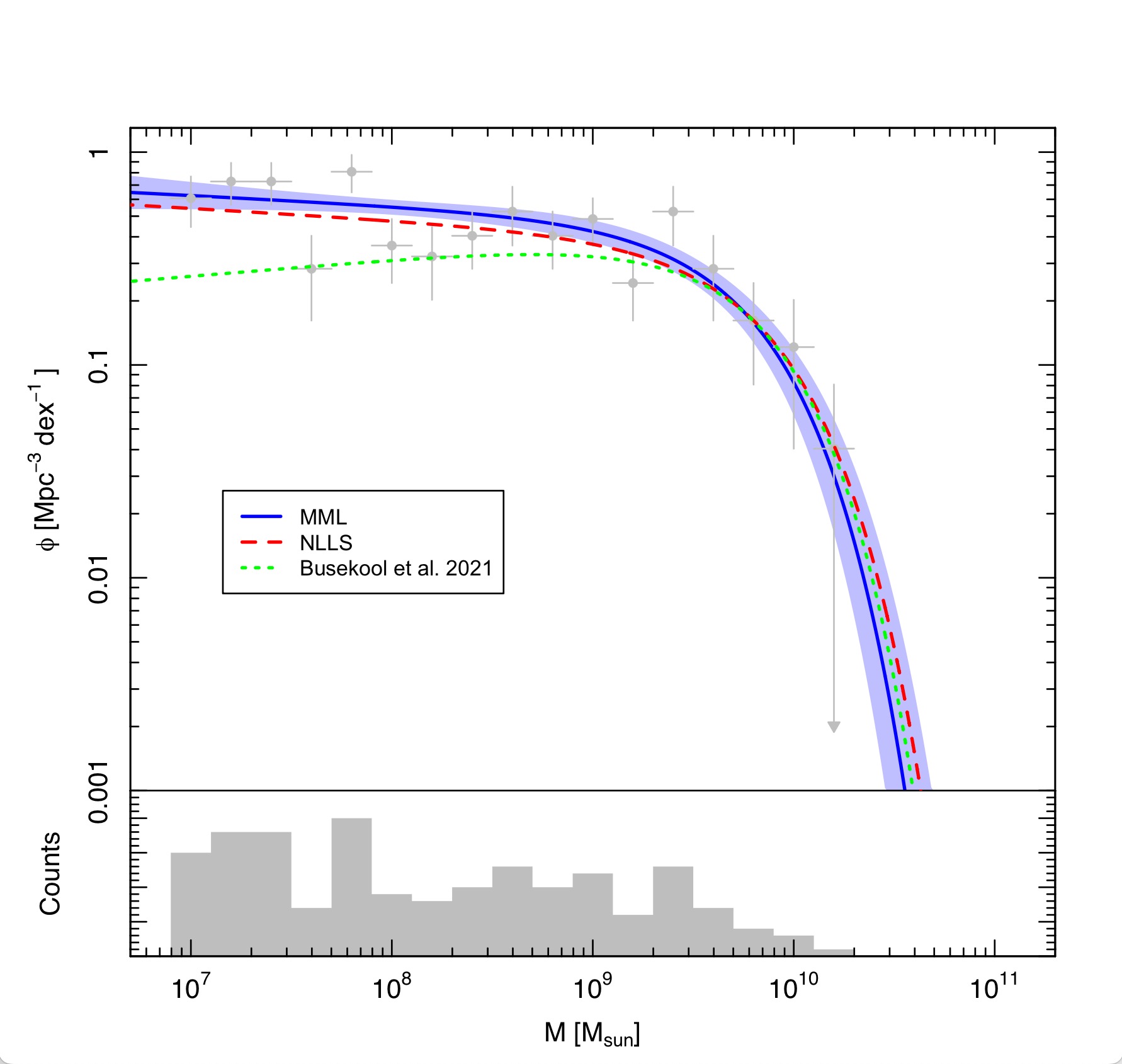}
    \caption{The top panel shows the H\,{\footnotesize I}MF of the UMa supergroup. 
    The grey points with error bars are the number densities within each bin. 
    The curve fitted by the MML method is the blue solid line, with the 1-$\sigma$ uncertainty in the blue area. 
    The red dashed line is the best fit from the LMFIT package based on the NLLS method.
    The result in \citet{Busekool2021} is shown with the green dotted line.
    The lower panel shows a histogram of the galaxy number distribution.}
    \label{fig:himf}
\end{figure}

\begin{figure}
	\includegraphics[width=\columnwidth]{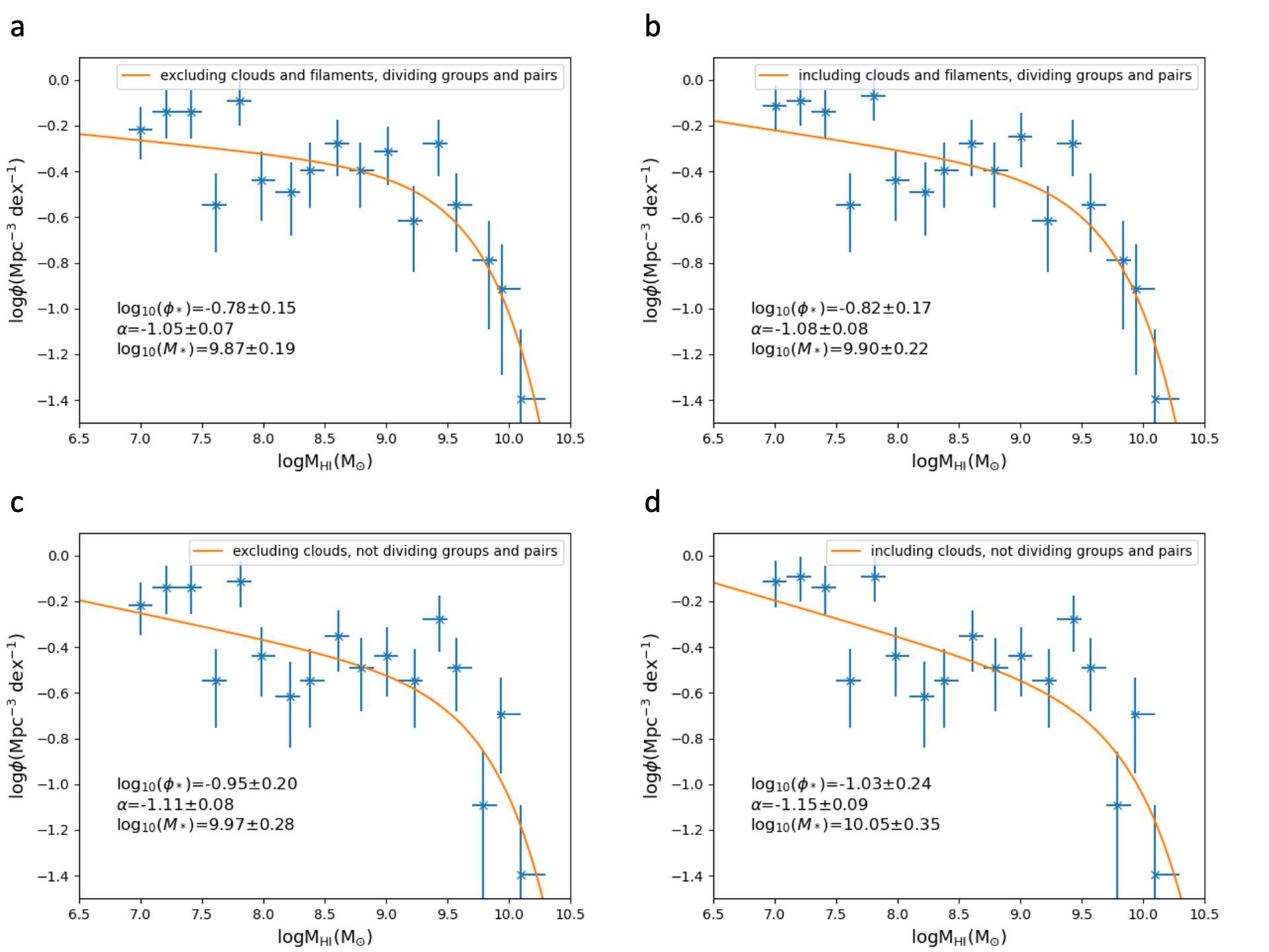}
    \caption{The H\,{\footnotesize I}MF calculated with the NLLS method in different cases: (a)excluding H\,{\footnotesize I} clouds and filaments, dividing small groups and confused pairs into individual galaxies; (b)including H\,{\footnotesize I} clouds and filaments, dividing small groups and confused pairs into individual galaxies; (c)excluding H\,{\footnotesize I} clouds, not dividing small groups and confused pairs into individual galaxies; (d)including H\,{\footnotesize I} clouds, not dividing small groups and confused pairs into individual galaxies.}
    \label{fig:himf_compare}
\end{figure}

The results of the fit are shown in Figure \ref{fig:himf}, with the top panel showing the H\,{\footnotesize I}MF fitted by the NLLS and MML method, and the bottom panel showing a histogram of the galaxy number distribution. 
The best-fit parameters based on the NLLS method are log$_{10}$($\phi_*$/Mpc$^{-3}$) = -0.78 $\pm$ 0.15, $\alpha$ = -1.05 $\pm$ 0.07 and log$_{10}$($M_*$/$M_{\sun}$) = 9.87 $\pm$ 0.19, and log$_{10}$($\phi_*$/Mpc$^{-3}$) = -0.70 $\pm$ 0.11, $\alpha$ = -1.05 $\pm$ 0.05 and log$_{10}$($M_*$/$M_{\sun}$) = 9.77 $\pm$ 0.13 for the MML method. 
The excluded H\,{\footnotesize I} clouds and filaments could be gas-rich ultra-faint dwarf galaxies whose optical signatures are not detected. 
Their exclusion in this case biases the H\,{\footnotesize I}MF calculations. 
In addition to this, the division of small groups and confused pairs into individual galaxies also biases the H\,{\footnotesize I}MF calculations, due to the larger errors in the H\,{\footnotesize I} masses of individual galaxies.
In comparison to the case of excluding H\,{\footnotesize I} clouds and dividing small groups and confused pairs, we made three comparison cases and the results are shown in Figure \ref{fig:himf_compare}. 
The results show that excluding H\,{\footnotesize I} clouds and filaments as well as dividing small groups and confused pairs flatten the slope at the low mass end of the H\,{\footnotesize I}MF and increase the characteristic mass at the high mass end. 
However, as they are small in percentage terms, the effect on the H\,{\footnotesize I}MF is insignificant and the parameter changes are roughly within the weighted fitting error.

The slope of the low mass end of our H\,{\footnotesize I}MF is steeper than the one derived from the VLA blind survey ($\alpha$ = -0.92 $\pm$ 0.16 and log$_{10}$($M_*$/$M_{\sun}$) = 9.8 $\pm$ 0.8), but not as steep as the H\,{\footnotesize I}MF's derived from HIPASS and ALFALFA.
The characteristic mass is more consistent.
This result is due to a large number of new FUMaS detections of galaxies with masses concentrated at 10$^7$-10$^8$ $M_{\sun}$, raising the slope at the low-mass end. 
Our finding of a flatter slope compared to the global H\,{\footnotesize I}MFs is in agreement with the conclusions of several investigators that the low mass slope is flatter in denser galaxy environments \citep{Rosenberg2002, Springob2005, Pisano2011, Westmeier2017, Said2019, Jones2020}, suggesting that the environment affects the H\,{\footnotesize I}MF. 
This is also consistent with theoretical predictions that galaxies in high-density regions are more likely to interact with each other, thus H\,{\footnotesize I} gas is stripped away and fewer low-mass galaxies are detected.
Note that the slope of our H\,{\footnotesize I}MF is almost identical to that of the optical LF found by \cite{Trentham2001},  which suggests that the gas stripping process is dominated by tidal stripping rather than ram pressure in the UMa supergroup, as ram pressure effect should produce more gas poor dwarf galaxies and resulting in different slopes between the LF and H\,{\footnotesize I}MF\citep{Trentham2001, Busekool2021}.
This result is also consistent with the fact that the UMa supergroup does not have a hot X-ray halo thus ram pressure stripping by hot gas is negligible.  

\section{Conclusions}
\label{sec:conclusion}

The UMa supergroup is an ideal target for studying the evolution of dwarf galaxies, galaxy interactions, and environmental influences on the H\,{\footnotesize I}MF, but it has not been explored thoroughly and lacks a complete H\,{\footnotesize I} source catalog. 
The FAST H\,{\footnotesize I} survey of the UMa supergroup is our first attempt to conduct a complete blind H\,{\footnotesize I} survey of the UMa region.
The main results are summarized as follows:

1. Using SoFiA for source finding and combining this with a manual search, we have obtained the most complete H\,{\footnotesize I} source catalog for the UMa supergroup.
This catalog contains 178 H\,{\footnotesize I} sources with velocities in the range 625-1213.4 km s$^{-1}$ and masses in the range 10$^{6.0}$-10$^{10.1}$ M$_{\sun}$ assuming a unity distance of 17.4 Mpc.
Compared to the previous UMa catalogs, we have detected all the H\,{\footnotesize I} sources previously reported in the literature,  and also detect H\,{\footnotesize I} signals for the vast majority of galaxies in the optical catalog (150 out of 166 galaxies in \citet{Pak2014}).
Compared to previous H\,{\footnotesize I} observations, FUMaS's fluxes are systematically high, with a median difference of 17.9\%.
This may be due to the fact that most of the H\,{\footnotesize I} sources in the UMa supergroup are extended sources, and the high sensitivity of FAST can detect more diffuse H\,{\footnotesize I} around extended sources. Such finding is consistent with that of \citet{Wang2024} based on FAST observations of the FEASTS sample.

2. There are 55 H\,{\footnotesize I} sources in the catalog that are detected for the first time.
In addition, 32 detections have no optical redshifts, for which 25 of them have corresponding galaxies in optical images, while the remaining 7 are most likely H\,{\footnotesize I} clouds without optical counterparts. 
The high sensitivity of FAST can provide a good complement to optical observations, adding redshift information for dwarf galaxies, and thereby increasing the number of known supergroup members.

3. There are many interacting systems in the UMa supergroup, and our catalog contains three confused pairs and three small groups.
We describe the four interacting systems in details. 
NGC 3992 and its three companion galaxies have overlapping gas disks; there is a filament of ~142 kpc between NGC 4026 and SDSS J115950.81+502955.3, and a filament of ~152 kpc extending south of NGC 4111; and the asymmetric gas disk of NGC 3998.

    4. We computed the H\,{\footnotesize I}MF of the UMa supergroup using the 1/V$_\mathrm{max}$ method and fitted it with both the NLLS and MML methods. 
    The best-fit parameters are log$_{10}$($\phi_*$/Mpc$^{-3}$) = -0.78 $\pm$ 0.15, $\alpha$ = -1.05 $\pm$ 0.07 and log$_{10}$($M_*$/$M_{\sun}$) = 9.87 $\pm$ 0.19 for the NLLS method, and log$_{10}$($\phi_*$/Mpc$^{-3}$) = -0.70 $\pm$ 0.11, $\alpha$ = -1.05 $\pm$ 0.05 and log$_{10}$($M_*$/$M_{\sun}$) = 9.77 $\pm$ 0.13 for the MML method.
    Compared to the VLA H\,{\footnotesize I}MF, the characteristic mass is consistent at the high-mass end, but the slope is steeper at the low-mass end because we detected more low-mass galaxies.
    Compared to the global H\,{\footnotesize I}MF, we find a flatter slope at the low-mass end, suggesting that gas stripping due to interactions in high-density galaxy regions reduces the number density of low-mass galaxies. 
    Such scenario is consistent with theoretical predictions for galaxy evolutions in dense environments.

\begin{acknowledgments}
We thank the anonymous referee for constructive suggestions which help to improve this work.
This work is supported by the Guizhou Provincial Science and Technology Projects (QKHFQ[2023]003, QKHPTRC-ZDSYS[2023]003, QKHFQ[2024]001-1).
FAST is a Chinese national mega-science facility, operated by the National Astronomical Observatories of Chinese Academy of Sciences (NAOC).
This work is also supported by the National Natural Science Foun-
dation of China (Grant Nos. 12373001).
\end{acknowledgments}

\bibliography{UMa}{}
\bibliographystyle{aasjournal}



\end{document}